\documentclass[notitlepage,pre,11pt,showkeys,showpacs,nofootinbib,tightenlines]{revtex4}
\usepackage{graphicx}
\graphicspath{{figs/}}
\usepackage{amsmath,amssymb}
\usepackage{bm}
\numberwithin{equation}{section}

\newcommand{\etal}{\textit{et~al.}}

\newcommand{\mathnotation}[2]{\newcommand{#1}{\ensuremath{#2}}}

\newcommand{\grad}{\nabla}
\newcommand{\lapl}{\Delta}
\renewcommand{\l}{\left} % \left
\renewcommand{\r}{\right} % \right
\mathnotation{\ldef}{\mathrel{\raisebox{.069ex}{:}\!\!=}}% Left define
\mathnotation{\rdef}{\mathrel{=\!\!\raisebox{.069ex}{:}}}% Right define
\newcommand {\degree} {^{\circ}}
\newcommand{\Perm}{\rm{Pe}}
\newcommand{\Pe}{\mathrm{Pe}}
\newcommand{\Pec}{{{\mathcal{P}}e}}
\mathnotation{\ks}{k_s}
\mathnotation{\ku}{k_u}
\mathnotation{\dint}{\mathrm{d}}
\newcommand{\fracs}[2]{#1/#2}

\begin{document}

\title{Stirring up trouble:\\
Multi-scale mixing measures for steady scalar sources\\
}
\author{Tiffany A. Shaw}
\email{tshaw@physics.utoronto.ca}
\affiliation{Department of Physics, University of Toronto,
Toronto, M5S 1A7, Canada}
\author{Jean-Luc Thiffeault}
\email{jeanluc@imperial.ac.uk}
\affiliation{Department of Mathematics, Imperial College
  London, SW7 2AZ, United Kingdom}
\author{Charles R. Doering}
\email{doering@umich.edu}
\affiliation{Department of Mathematics and Michigan Center
for Theoretical Physics\\
University of Michigan, Ann Arbor, MI 48109-1043, USA}

\begin{abstract}
The mixing efficiency of a flow advecting a passive scalar sustained by steady
sources and sinks is naturally defined in terms of the suppression of
bulk scalar variance in the presence of stirring, relative to the
variance in the absence of stirring.  These variances can be weighted at
various spatial scales, leading to a family of multi-scale mixing measures and
efficiencies.  We derive {\it a priori} estimates on these efficiencies from
the advection--diffusion partial differential equation, focusing on a broad
class of statistically homogeneous and isotropic incompressible flows.  The
analysis produces bounds on the mixing efficiencies in terms of the P\'eclet
number, a measure the strength of the stirring relative to molecular
diffusion.  We show by example that the estimates are sharp for particular
source, sink and flow combinations.  In general the high-P\'eclet number
behavior of the bounds (scaling exponents as well as prefactors) depends on
the structure and smoothness properties of, and length scales in, the scalar
source and sink distribution.  The fundamental model of the stirring of a
monochromatic source/sink combination by the random sine flow is investigated
in detail via direct numerical simulation and analysis.  The large-scale
mixing efficiency follows the upper bound scaling (within a logarithm) at high
P\'eclet number but the intermediate and small-scale efficiencies are
qualitatively less than optimal.  The P\'eclet number scaling exponents of the
efficiencies observed in the simulations are deduced theoretically from the
asymptotic solution of an internal layer problem arising in a quasi-static
model.
\end{abstract}

\keywords{stirring, mixing, advection, diffusion, eddy diffusion, turbulent diffusion}
\pacs{47.27.Qb, 92.10.Lq, 92.60.Ek, 94.10.Lf }

\maketitle

\section{Introduction}
\label{sec:intro}

Mixing processes in fluids play a key role in a wide variety of engineering
applications and for natural systems such as the ocean and atmosphere.  Their
theoretical study has been a major focus of research, as indicated by the
large number of review articles~\cite{Ottino1990,Majda1999,Warhaft2000,%
Shraiman2000,Sawford2001,Falkovich2001,Aref2002,Wiggins2004}.  At the smallest
scales mixing is achieved by molecular diffusion processes, but it may be
facilitated greatly by stirring.  The result of stirring is usually to enhance
the effect of molecular diffusion and increase the mixing
rate~\cite{Batchelor1949,Batchelor1952a,Batchelor1959,Kraichnan1968,%
Kraichnan1974}.  Quantitative understanding of the fundamental features of
stirring and its influence on mixing processes is important for the effective
modeling, simulation and design or control of these systems.

The ``efficiency'' of mixing means different things in different contexts.
For example the dispersion of an initial distribution by an imposed flow is a
transient problem where the temporal approach to the final fully mixed state,
rather than the final state itself, is of central interest.  Consider for
definiteness the homogeneous advection--diffusion equation for a passive
scalar field $\theta(\bm{x},t)$ stirred by a divergence-free velocity field
$\bm{u}(\bm{x},t)$,
\begin{equation}
  \frac{\partial \theta}{\partial t} + \bm{u} \cdot \nabla \theta = \kappa
  \lapl \theta
  \label{eq:ADhomo}
\end{equation}
where $\kappa$ is the molecular diffusivity.  If this equation is supplied
with initial concentration $\theta(\bm{x},0)$ and applied in an appropriate
domain without sources, sinks or scalar flux at the boundaries, then the
integral of $\theta$ is conserved, so without loss of generality it may be
taken to vanish from the start.  But the $L_{2}$-norm
$\|\theta(\cdot,t)\|_{2}$, proportional to the scalar variance in finite
volume domains, decreases with time.  Indeed, multiplying (\ref{eq:ADhomo}) by
$\theta$ and integrating by parts,
\begin{equation}
  \frac{d}{dt}  \|\theta\|_{2}^{2} = - 2 \kappa\|\nabla \theta\|_{2}^{2},
  \label{eq:ADhomodecay}
\end{equation}
indicating an inexorable decay of the variance.  Efficient mixing in this
transient decay problem means faster decay of the scalar variance.  The mixing
efficiency of a particular flow could be defined, for example, in terms of its
ability to reduce the variance from the initial value to a prescribed value
within a specific period of time~\cite{Constantin2005}.  Because the
right-hand side of (\ref{eq:ADhomodecay}) is proportional to $\kappa$, it is
evident that molecular processes are ultimately responsible for mixing by this
criterion.  Even though the stirring field does not appear explicitly in
(\ref{eq:ADhomodecay}), the conventional intuition is that material line
stretching in the flow can amplify scalar gradients thereby enhancing the
molecular mixing rate.  Indeed, the velocity's rate-of-strain matrix serves as
the local growth rate of the scalar gradient field. These issues are of extreme interest for both theory and
applications, but in this paper we are interested in a distinct scenario where
different effects are at work.  

Mixing a scalar field whose fluctuations are constantly replenished by steady
but spatially inhomogeneous sources and sinks is a problem with a long
history.  Early on, Townsend~\cite{Townsend1951,Townsend1954} was concerned
with the effect of turbulence and molecular diffusion on a line-source of
temperature, a heated filament.  The spatial localization of the source,
imposed by experimental constraints, enhanced the role of molecular
diffusivity.  Saffman~\cite{Saffman1960} also found that molecular diffusion
and turbulent diffusion were not simply additive and that higher-order
corrections were needed.  Durbin~\cite{Durbin1980} and
Drummond~\cite{Drummond1982} introduced stochastic particle models to
turbulence modeling, and these allowed more detailed studies of the effect of
the source on diffusion.  Sawford and Hunt~\cite{Sawford1986} pointed out that
small sources, such as heated filaments, lead to an explicit dependence of the
variance on molecular diffusivity.  Many refinements to these models followed,
see for instance~\cite{Thomson1990,Borgas1994} and the review by
Sawford~\cite{Sawford2001}.  Chertkov
\etal~\cite{Chertkov1995,Chertkov1995b,Chertkov1997,Chertkov1997b,%
Chertkov1998} and Balkovsky \& Fouxon~\cite{Balkovsky1999} treated the case of
a random, statistically-steady source.  Our goal in the present paper is to
make the source-dependence of the concentration variance more precise by
working directly from the advection--diffusion equation, without specifying
the underlying turbulent statistics other than basic stationarity and
homogeneity assumptions.

%\section{Advection--Diffusion with a Source}
%\label{eq:ADS}

When a source of scalar concentration is present, the transient kinetics are
of less immediate interest and instead the properties of the (statistical)
steady state are of greater relevance.  As will be seen, this sustained
steady-state dynamics highlights other features of stirring and mixing
processes.  In comparing the steady-state problem to the transient problem
defined by Eq.~\eqref{eq:ADhomodecay}, it is important to remember that the
long-time asymptotic behavior of the decaying problem is usually irrelevant
to the corresponding long-time behavior of the steady-state problem.  This is
because the continuous replenishing of concentration overwhelms
small-amplitude effects observed for long-time decay, such as the `strange
eigenmode'~\cite{%
Pierrehumbert1994,Antonsen1996,Rothstein1999,Fereday2002,Sukhatme2002,%
Wonhas2002,Pikovsky2003,Thiffeault2003d,Thiffeault2004b,Schekochihin2004,%
Vanneste2005,Gilbert2006}.

In this paper we consider the stirring and mixing of a passive scalar
sustained by a steady source-sink function~$s(\bm{x})$.  Given a prescribed
divergence-free velocity field $\bm{u}(\bm{x},t)$ and a molecular
diffusivity~$\kappa$, the scalar concentration $\theta(\bm{x},t)$ obeys the
inhomogeneous advection--diffusion equation
\begin{equation}
  \frac{\partial \theta}{\partial t} + \bm{u} \cdot \nabla \theta =
  \kappa \lapl \theta + s(\bm{x})
  \label{eq:AD}
\end{equation}
supplemented with initial concentration field $\theta(\bm{x},0)$.  We consider
a domain without any net scalar flux at the boundaries: the
periodic box of size $L$, i.e., $\bm{x} \in \mathbb{T}^d$, the $d$-dimensional
torus of volume $L^{d}$.  The spatial mean of $\theta$ is computed
immediately,
\begin{equation}
   \frac{1}{L^{d}} \int \theta(\bm{x},t) \, \dint^{d}x \ = \
   \frac{1}{L^{d}} \int \theta(\bm{x},0) \, \dint^{d}x \ + \
   t \times \frac{1}{L^{d}} \int s(\bm{x}) \, \dint^{d}x \, ,
\end{equation}
and deviations from the spatial mean satisfy (\ref{eq:AD}) with
$s(\bm{x})$ replaced by $s(\bm{x}) - L^{-d} \int s \, \dint^{d}y$.  So to study
the fluctuations we may assume without loss of generality that
$\theta(\bm{x},0)$ and $s(\bm{x})$, and thus also
$\theta(\bm{x},t)$ have spatial mean zero.

Fluctuations in the scalar concentration are natually measured in terms of the
steady-state variance $\langle \theta^2 \rangle$, where we introduce the
space-time average.  The two averaging operations we use are the time average
\begin{equation}
  \overline{F}(\bm{x})\ldef\lim_{t \rightarrow \infty}
  \frac{1}{t} \int_0^t F(\bm{x},t') \, dt',
\end{equation}
assuming as necessary that the limit exists, and the space-time average
\begin{equation}
  \langle F \rangle \ldef \frac{1}{L^{d}} \int \overline{F}(\bm{x})
  \, \dint^{d}x \, .
\end{equation}
Effective stirring makes the scalar field more spatially uniform, lowering the
variance, and this is the basic mixing effect that we set out to study.  Many
investigations have been concerned with other statistical properties of the
scalar field for this kind of model, such as details of the tails of the
probability distribution of $\theta$~\cite{Majda1999,Falkovich2001}.  While
these studies present fascinating mathematical and physical issues, in terms
of applications they are most likely to be of ultimate use in designing
closure approximations, i.e., models of the model, in order to accurately
estimate bulk measures of mixing like variance reduction.  In this work we
focus directly on the supression of the scalar fluctuations as a primary
indicator of mixing.

In terms of the Fourier decomposition of the scalar field,
\begin{equation}
  \hat{\theta}_{\bm{k}}(t) = \frac{1}{L^d}
  \int \theta({\bm{x}},t)\,e^{-i\bm{k}\cdot \bm{x}}\,
  \dint^{d}x
\end{equation}
where ${\bm{k}}=(2\pi/L){\bm{n}}$ for ${\bm{n}}=(n_1,\ldots,n_d)
\in \mathbb{Z}^d$, the steady-state variance is
\begin{equation}
   \langle \theta^2 \rangle = \sum_{\bm{k}}
   \overline{|\hat{\theta}_{\bm{k}}|^{2}}.
\end{equation}
The magnitude $k$ of the wavenumbers naturally index spatial scales, so
information about the fluctuations at different scales may be obtained
by weighting the sums of the Fourier coefficients~\cite{Mathew2005}.  The
simplest indicators of the fluctuations on small and large scales are,
respectively,
\begin{equation}
   \langle |\nabla \theta|^2 \rangle = \sum_{\bm{k}} k^{2}
   \overline{|\hat{\theta}_{\bm{k}}|^{2}}
\end{equation}
and 
\begin{equation}
   \langle |\nabla^{-1} \theta|^2 \rangle = \sum_{\bm{k}} 
   \overline{|\hat{\theta}_{\bm{k}}|^{2}}/k^{2}
\end{equation}
where the inverse gradient $\nabla^{-1}$ is defined in Fourier space as
multiplication by $- i {\bm k}/k^{2}$, a well-defined operator on these
functions with spatial mean zero.  For the purposes of this study the
effectiveness of stirring on mixing at relatively small and large scales
will be gauged in terms of these norms.  Efficient stirring decreases the
variances on all scales, although it can be expected that any particular
stirring may be more effective on some scales than on others.

It is appropriate here to point out a fundamental and elementary distinction
between transient and steady-state stirring.  For efficient transient mixing
the goal is to decrease the scalar variance $\sim \|\theta(\cdot,t)\|_{2}^{2}$
as quickly as possible by {\it increasing} the gradient variance $\sim
\|\nabla \theta(\cdot,t)\|_{2}^{2}$ via stirring.  However in the steady-state
problem stirring can only {\it reduce} the mean scalar gradient variance---and
thus the mean rate of variance decay---from its purely diffusive value in the
absence of stirring.

The proof of this (perhaps unexpected) fact is easy.
Multiplying (\ref{eq:AD}) by $\theta$ and averaging over space and time
with appropriate integrations by parts produces the well-known balance
\begin{equation}
   \kappa \langle |\nabla \theta|^2 \rangle = 
   \langle \theta s \rangle.
   \label{elphaba}
\end{equation}
Inserting the gradient and its inverse on the right-hand side, integrating by
parts again, and then employing the Cauchy--Schwarz inequality yields
\begin{equation}
   \kappa \langle |\nabla \theta|^2 \rangle \ = \
   \langle \theta  \nabla \cdot \nabla^{-1}s \rangle \ = \
   - \langle \nabla \theta \cdot \nabla^{-1}s \rangle \ \le \
   \langle |\nabla \theta|^{2}\rangle^{1/2}
   \langle |\nabla^{-1}s| \rangle^{1/2}.
   \label{glinda}
\end{equation}

Note that the long-time solution of (\ref{eq:AD}) in the absence of stirring
is the steady-state solution $\theta_{0}(\bm{x})$ of the diffusion equation
with source $s(\bm{x})$,
\begin{equation}
  \theta_{0} \ = \ - \frac{1}{\kappa} \lapl^{-1} s \, ,
  \label{zero}
\end{equation}
where the inverse Laplacian is multiplication by $-k^{-2}$ in Fourier space.
Solving for the steady-state scalar gradient variance in (\ref{glinda}) and
noting that $\nabla \theta_{0} = -\kappa^{-1} \nabla^{-1} s$, we conclude that
\begin{equation}
  \langle |\nabla \theta|^2 \rangle \le 
  \langle |\nabla \theta_{0}|^2 \rangle.
  \label{boc}
\end{equation}

This relationship is uniform in the advecting velocity field
$\bm{u}(\bm{x},t)$ implying that no clever stirring on any scales can increase
the mean scalar gradient variance over its baseline unstirred value.  In light
of this observation we may anticipate that stirring strategies designed to
maximize mixing efficiency in the sustained source problem are likely to be
different from those employed in the transient decay scenario.  We remark that
this result holds in the absence of boundaries; the~$\kappa\rightarrow 0$
limit could be much more complicated in the presence of boundary layers.

The variances $\langle |\nabla^{p} \theta_{0}|^2 \rangle$, defined above for
$p \in \{-1,0,1\}$, are both dimensional and dimensionally distinct quantities.
In order to compare them with each other or to compare different
physical systems we need sensible nondimensional measures.
Hence we define the dimensionless {\it multi-scale mixing efficiencies}, denoted
${\cal E}_{p}$ for $p \in \{-1,0,1\}$, via
\begin{equation}
  {\cal E}_{p}^{2} = \frac {\langle |\nabla^{p} \theta_{0}|^2 \rangle}
  {\langle |\nabla^{p} \theta|^2 \rangle}
\end{equation}
where $\theta_{0}$ is the steady solution of the unstirred problem defined in
(\ref{zero}).  Effective stirring decreases scalar variances relative to those
due to diffusion alone, increasing these mixing efficiences.  The calculation
resulting in (\ref{boc}) has established that ${\cal E}_{1} \ge 1$.

Intensifying the stirring often increases the mixing efficiencies and it is
important to characterize this property in terms of the forcefulness of the
flow.  The simplest bulk measure of the vigor of the velocity field is its
mean kinetic energy, or equivalently the {\it rms} speed $U$ defined by
\begin{equation}
  U^{2} = \langle \lvert\bm{u}\rvert^2 \rangle.
\end{equation}
The nondimensional measure of the strength of the stirring relative to the
effect of molecular viscosity is the P\'eclet number $\Pe$ that we define as
\begin{equation}
  \Pe = \frac{UL}{\kappa}
\end{equation}
using the domain length scale $L$ for simplicity here.  It will become
apparent that this domain length scale may not be the most appropriate one for
this purpose; determining the relevant length scales is one of the central
points of this study.

For many applications it is useful to know how fluctuations at various length
scales may be suppressed as functions of $\kappa$, $U$, and other features of
the problem such as details of the flow and source-sink structures.  Toward
this end it is desirable to know how the multi-scale mixing efficiencies depend
on the P\'eclet number, and the notion of ``eddy diffusivity'' provides a
conceptual benchmark for this dependence.

A flow with velocity scale $U$ and ``persistence length'', ``mixing
length'' or ``eddy size'' $\ell$ that characterizes the typical
distance a particle travels before changing direction can disperse
particles diffusively on appropriate space and/or time scales.  This
suggests that when advection dominates molecular diffusion, an
effective diffusion with coefficient $\kappa_{\mathrm{eff}} \sim
U\ell$ might replace the advection to determine some gross statistical
features of the scalar field.\footnote{Indeed, the situation where
  $\ell$ is much smaller than any length scales in the initial data
  $\theta(\bm{x},0)$ or the source $s(\bm{x})$ is the setting for
  homogenization
  theory~\cite{Majda1999,MR1994746,MR2052862,MR2142878}.}  If this is
so, then the steady-state scalar variances are all $\sim
\kappa_{\mathrm{eff}}^{-2}$ and the efficiencies are all $\sim
\kappa_{\mathrm{eff}}/\kappa$.  According to this argument
\begin{equation}
   {\cal E}_{p} \sim \frac{U\ell}{\kappa} = \frac{\ell}{L} \times \Pe.
\end{equation}

The linear scaling ${\cal E}_{p} \sim \Pe$ at high P\'eclet numbers, which we
will refer to as the ``classical'' scaling, provides a baseline reference for
the multi-scale mixing efficiencies.  Flows that generate this classical
scaling asymptotically as $\Pe \rightarrow \infty$ produce a truly effective
``residual'' molecular-like diffusion as far as the suppresion of variances at
the various spatial scales is concerned, even in the singular $\kappa
\rightarrow 0$ limit of vanishing molecular diffusivity.  And as this
discussion suggests, if the efficiency scales classically then the prefactor
provides a precise prediction of the length scale with which an eddy
diffusivity might be meaningfully identified.

The principal purpose of this paper is to determine limits on the mixing
efficiencies, i.e., to derive {\it a priori} bounds on ${\cal E}_{p}$ as a
function of $\Pe$, and to investigate what sort of flows might realize those
limits.  Upper bounds on ${\cal E}_{p}(\Pe)$ are of particular interest
because they characterize the most efficient stirring strategies that can
possibly be hoped for.  If the bounds are to be useful then they should be
realizable or approachable, or at the very least indicative of the kind of
behavior that is possible---such as a scaling like ${\cal E}_{p} \sim
\Pe^{\alpha}$ at high P\'eclet numbers.

Upper bounds on the mixing efficiencies follow from lower bounds on
the variances and the first study in this direction was apparently by
Thiffeault, Doering, and Gibbon \cite{Thiffeault2004} who focused on
estimates and simulations for ${\cal E}_{0}$.  (Related bounds on heat
kernels, with and without flow, have been known for a long
time~\cite{MR0100158,MR1085646,MR1246475,MR1482931,MR2031461,%
Davies1993,Kusoda1988}.)  They
adapted an approach that had been used to bound turbulent dissipation
in the Navier--Stokes equations \cite{Doering2002,Doering2003} for
application to inhomogeneous advection--diffusion equations.  For the
steady source model of interest here they showed that if $\langle
\theta_{0}^{2} \rangle < \infty$ then ${\cal E}_{0} \le a + b\, \Pe$
where the coefficients $a$ and $b$ are homogeneous scale invariant
functionals of the source $s$, i.e., invariant under the
transformation $s({\bm x}) \rightarrow c\, s({c' \bm x})$ for any
constants $c$, and $c' \ne 0$.

That result showed that very generally the classical scaling is an upper limit
to the mixing efficiency in this most basic sense.  Moreover, the coefficient
in the high P\'eclet number scaling ${\cal E}_{0} \lesssim b \, \Pe$ puts a
limit on any reasonable value for a mixing length: $\ell \lesssim b L$.  It is
especially notable that this rigorous estimate of $\ell$ (really of the
prefactor $b$) is uniform in the stirring field $\bm{u}$ and independent of
any length scales it exhibits.  It also is independent of $\kappa$.  It
emerges as a length scale in the source-sink distribution, which is seen to
play a more important role in the mixing process than the conventional eddy
diffusion picture anticipates.  When the relevant length scale in the
sustaining source and sink is small then the variance suppression by advection
is necessarily limited by this no matter what spectrum of scales is present in
the stirring process.

These observations serve as the starting point for this study.  Here we carry
forward the investigation of stirring and variance suppression by extending
the analysis to multi-scale mixing measures while focusing on a broad but
specific class of statistically stationary homogeneous and isotropic flows.
The restriction to this class of flows---a class that includes but is not
limited to homogeneous isotropic high Reynolds number turbulence---allows
for the exact solution of some variational problems for bounds on the mixing
efficiencies.  For certain sources and sinks these bounds yield anomalous
sub-classical exponents for the $\Pe$ scaling of some of the ${\cal E}_{p}$.
Thus anomalous scaling is inevitable for some source-sink distributions.
In those cases it cannot be avoided by manipulating details of the flow or
the spectrum of the stirring.
We study the case of a monochromatic source 
$s({\bm x}) \sim \sin({\bm k}_{s} \cdot {\bm x})$ in detail.  
The estimates are particularly simple in this case and they are
sharp: we will exhibit a statistically stationary homogeneous and
isotropic stirring strategy that saturates the upper bounds.

The rest of this paper is organized as follows.  We introduce the relevant
class of statistically stationary, homogeneous and isotropic flows in
Section~\ref{sec:SHIF}, and present some specific examples.  We formulate
variational problems for bounds on the mixing efficiencies in
Section~\ref{sec:BOUNDS}, and derive general estimates.  In
Section~\ref{sec:saturation} we evaluate the bounds explicitly for some
particular sources and sinks and show that classical scaling estimates may
be sharp.  In Section~\ref{sec:SARSAS} we focus on the high-$\Pe$ behavior of
the mixing efficiency bounds for a variety of source-sink distributions,
and show that anomalous sub-classical scaling is sometimes unavoidable.

In Section~\ref{sec:SSSSSF} we are concerned with the fundamental example of a
monochromatic source stirred by a flow with a single length scale, the
so-called random sine flow, a type of renewing flow.  Measuring the mixing
efficiencies in direct numerical simulations, we find that the large-scale
efficiency ${\cal E}_{-1}$ scales (nearly) classically with respect to $\Pe$,
like its upper bound, but the intermediate and small-scale efficiencies ${\cal
E}_{0}$ and ${\cal E}_{1}$ scale sub-classically, i.e., with powers of $\Pe$
less than 1.  We show that the anomalous exponents for ${\cal E}_{0}$ and
${\cal E}_{1}$ can be deduced from the asymptotic analysis of a static flow
problem.  The concluding Section~\ref{sec:SandD} contains a summary of the
results along with a discussion of open problems and compelling future
challenges.  Some technical details are relegated to appendices.

\section{Statistically Stationary Homogeneous Isotropic Flows}
\label{sec:SHIF}

Our approach to estimating mixing efficiencies is kinematic: the stirring
vector field ${\bm u}({\bm x},t)$ is assumed to be given.  It could be a
solution of the Navier--Stokes equations, or it could be a stochastic process
with convenient or interesting spectral properties, or it could be a regular
time-periodic field.  The mixing efficiency bounds obtained in this paper will
apply so long as a few generic statistical conditions are satisfied.  Of
course not every stirring field satisfying these conditions will saturate the
bounds, but they are all limited by them.

This analysis in this paper is concerned with velocity fields that satisfy the
following three conditions:
\begin{itemize}
\item{The field is divergence free, $\nabla \cdot {\bm u} = 0$,
everywhere and at all times.
%We want to consider incompressible volume-preserving flows.
}
\item{The field has finite mean kinetic energy, $U^{2} = \langle |{\bm u}|^{2}
\rangle < \infty$, so that the P\'eclet number $\Pe = UL/\kappa$ is finite.
We will also presume more regularity for the velocity whenever necessary to
carry out formal calculations.  This will be apparent in the course of the
applications if the ultimate estimates depend on other norms of the field.}
\item{The velocity field is statistically stationary, homogeneous and
isotropic.  For the purposes of this work these qualities are defined by the
one- and two-point equal time statistics (presuming that these time averages
exist pointwise in space)
\begin{equation}
 \overline{u_i(\bm{x},\cdot)} = 0
 \label{mean1}
\end{equation}
\begin{equation}
  \overline{u_i(\bm{x},\cdot)u_j(\bm{y},\cdot)} \ = \ C_{ij}(\bm{x}-\bm{y})
  \ = \ \sum_{\bm{k} \ne 0} \frac{\hat{C}(k)}{d-1}
  \ \l(\delta_{ij} - \frac{k_{i}k_{j}}{k^{2}}\r)
  \ e^{i\bm{k}\cdot(\bm{x}-\bm{y})} \ + \ \frac{\hat{C}(0)}{d}\delta_{ij}
  \label{eq:SHIT}
\end{equation}
where $\hat{C}(k)$ depends only on the magnitude $k$ of the wavenumber
$\bm{k}$.  }
\end{itemize}
The conditions of finite energy and incompressiblity are familiar and
straightforward.  In the remainder of this section we discuss these
stationarity, homogeneity and isotropy conditions and their relevant
implications for the calculation of bounds on the various multi-scale mixing
efficiencies.  We also provide explicit examples, i.e., we describe several
flows with these properties that will be useful for considerations
in subsequent sections.

First, setting $\bm{x} = \bm{y}$ in (\ref{eq:SHIT}) produces the single point
component-by-component covariance
\begin{equation}
  \overline{u_i(\bm{x},\cdot)u_j(\bm{x},\cdot)} \ = \ C_{ij}(0)
  \ = \ \sum_{\bm{k} \ne 0} \frac{\hat{C}(k)}{d-1}
  \ \l(\delta_{ij} - \frac{k_{i}k_{j}}{k^{2}}\r)
  \ + \ \frac{\hat{C}(0)}{d}\delta_{ij}.
\end{equation}
Note that because $\hat{C}(k)$ depends only on the magnitude of the wavenumber,
\begin{equation}
\sum_{\bm{k \ne 0}} \hat{C}(k) \ \frac{k_{i}k_{j}}{k^{2}}
 \ = \ \frac{1}{d} \sum_{\bm{k} \ne 0} \hat{C}(k) \delta_{ij}.
\end{equation}
Thus
\begin{equation}
  \overline{u_i(\bm{x},\cdot)u_j(\bm{x},\cdot)} \ = \ 
  \frac{1}{d} \, \sum_{\text{all} \ \bm{k}} \hat{C}(k) \ \delta_{ij} \ = \
  \frac{U^2}{d} \delta_{ij}
  \label{eq:HIT1}
\end{equation}
where we identify the mean square velocity $U^{2} = \langle |{\bm u}|^{2}
\rangle = \sum \hat{C}(k)$.

Then if the velocity field is sufficiently regular (or equivalently if
$\hat{C}(k)$ decays sufficiently fast as $k \rightarrow \infty$) we may also
deduce some correlations of the derivatives of ${\bm u}$.  For example
differentiating $C_{ij}(\bm{x}-\bm{y})$ in (\ref{eq:SHIT}) by $x_{k}$ and
setting $\bm{y}=\bm{x}$ leads to
\begin{equation}
 \overline{\frac{\partial u_i(\bm{x},\cdot)}
 {\partial x_k}u_i(\bm{x},\cdot)} \ = \ 
 \sum_{\bm{k \ne 0}} \frac{i \,\hat{C}(k)}{d-1}
  \ \l(\delta_{ij} k_{k} \,  - \frac{k_{i}k_{j} k_{k}}{k^{2}}\r)
 \ = \ 0
 \label{eq:HIT2}
\end{equation}
because $\hat{C}(k)$ summed against an odd number of orthogonal components
vanishes when the sum is absolutely convergent.  Differentiating
$C_{ij}(\bm{x}-\bm{y})$ by $x_{k}$ and $y_{l}$, setting $\bm{y}=\bm{x}$, and
subsequently contracting over $k$ and $l$ gives
\begin{equation}
   \overline{ \frac{\partial u_i(\bm{x},\cdot)}{\partial x_k}\frac{\partial u_j(\bm{x},\cdot)}
   {\partial x_k}} \ = \
   \frac{1}{d} \, \sum_{\bm{k} \ne 0} k^{2}\, \hat{C}(k) \, \delta_{ij}
   \ = \
   \frac{\Omega^2}{d} \delta_{ij}
   \label{eq:HIT3}
\end{equation}
where we identify the enstrophy 
$\Omega^{2} \ldef \langle |\nabla {\bm u}|^{2} \rangle = \sum k^{2}\hat{C}(k)$.
In the context of statistical turbulence theory the ratio
$\lambda = U/\Omega$ is (proportional to) the Taylor microscale. 

An example of such a statistically stationary homogeneous isotropic flow is
the solution ${\bm u}({\bm x},t)$ of the incompressible Navier-Stokes
equations
\begin{equation}
   \frac{\partial \bm u}{\partial t} + {\bm u} \cdot \nabla {\bm u} + \nabla p
   \ = \ \nu \Delta {\bm u} + {\bm f}({\bm x},t)\, , \quad 
   \nabla \cdot {\bm u} \ = \ 0,
     \label{eq:NS}
\end{equation}
where $\nu$ is the kinematic viscosity and ${\bm f}({\bm x},t)$ is a spatially
periodic body force applied to maintain a statistical steady state.  Of course
the forcing would have to be capable of producing the homogeneous and
isotropic statistics described above, not an altogether trivial task although
it is generally expected that suitably homogeneous and isotropic random forces
will achieve it.  This approach would be taken to study mixing properties of
high-Reynolds number turbulent flows.

Other flows may be easier and more convenient to implement in simulations or
to utilize in the analysis of specific models.  Some features of the
stirring---for instance a stationary energy spectrum consistent with developed
turbulence---can be realized by specifying the modal amplitudes $\hat{{\bm
u}}_{\bm{k}}(t)$ as mean-zero stochastic processes with $\bm{k} \cdot
\hat{{\bm u}}_{\bm{k}}(t) = 0$ and appropriately uncorrelated for distinct
wavenumbers, to produce any desired energy spectrum $\hat{C}(k) =
\overline{|\hat{{\bm u}}_{\bm{k}}|^{2}}$.  This is possible
when $\overline{|\hat{{\bm u}}_{\bm{k}}|^{2}}$ depends only on the magnitude
$k$ of the wavenumber $\bm{k}$. 

One model of interest is a flow involving just a single wavenumber $k \ne 0$,
the random sine flow, a.k.a.\ the renewing wave
flow~\cite{Pierrehumbert1994,Antonsen1996,Thiffeault2004,Tsang2005}.  (The
term `renewing' or `renovating' has been used to refer to flows that are
piecewise-constant in time but change randomly at regular
intervals~\cite{Dittrich1984,Zeldovich1984,Gilbert1992}.
Zeldovich~\cite{Zeldovich1982} introduced a similar single-mode flow, but it
was oscillatory rather than renewing and thus had poor mixing properties.)  In
this example the velocity vector field switches periodically among steady
shearing flows of the form
\begin{equation}
  \bm{u}(\bm{x}) \ = \ \sqrt{2} \, \bm{U} \,
  \sin{(\bm{k}\cdot\bm{x} + \varphi)}
   \label{eq:ZSF}
\end{equation}
where $\bm{k}\cdot\bm{U}=0$, $|\bm{U}| = U$ and the phase $\varphi$ is
selected independently and uniformly from $[0,2\pi)$ upon each switch.
The switching may be strictly periodic or random in time, and
in either case the characteristic persistence $\tau$
is another parameter of this flow.
Of course an appropriate selection of the directions for $\bm{k}$ and
$\bm{U}$ must be made in order to realize statistical homogeneity
and isotropy by the definition used here.

The kinetically simplest possible example of a statistically homogeneous and
isotropic flow is one where only $\hat{C}(0) \ne 0$.  At each instant of time
the flow is then spatially uniform, a steady wind $\bm{u}(\bm{x}) = \bm{U}$
with the direction switching periodically or randomly in time so that the time
average of $\bm{u}$ vanishes and the component-component correlation satisfies
$\overline{u_{i}u_{j}} \sim \delta_{ij}$.  This wind could sample many
directions, or as few as $2 \times d$ in the $\pm$ directions along an
orthogonal set of coordinate axes.

%===================================================

\section{Bounds on the Mixing Efficiencies}
\label{sec:BOUNDS}

In this section we derive bounds on the multi-scale mixing efficiencies ${\cal
E}_{p}$ for~$p=0,1,-1$, corresponding to intermediate, small, and large
scales, respectively, for flows that satisfy the statistical homogeneity and
isotropy conditions in \eqref{eq:HIT1} and \eqref{eq:HIT2}.  We first briefly
address lower bounds, but focus for the most part on upper estimates.  The
variational formulation and solution for upper bounds on intermediate and
large length scales proceeds along similar lines, so we shall treat the
variance in detail and give a more brisk derivations for the large-scale
mixing measure.  Two different lower estimates on the small-scale variance,
corresponding to upper bounds on the mixing efficiency at small scales, are
derived.  One of these small-scale results depends on the spectral
distribution of energy in the stirring field while the other depends only on
the total bulk energy.

\subsection{Lower bounds on the mixing efficiencies}
\label{sec:bound_{lower}}

\textbf{(See Appendix~\ref{sec:corr} for a corrigendum.)}

\medskip

Lower bounds on the mixing efficiencies follow from upper bounds on the
corresponding variances.  We already derived a lower bound for 
the small-scale mixing efficiency ${\cal E}_{1}$ in the introduction by
considering the
steady-state variance dissipation-production balance in (\ref{elphaba}),
\begin{equation}
\kappa \langle |\nabla \theta|^{2} \rangle = 
\langle s \theta \rangle.
\label{elphaba2}
\end{equation}
The subsequent result in (\ref{boc}), valid for any incompressible flow even
without invoking any statistical assumptions, is precisely the statement that
${\cal E}_{1} \ge 1$.

We expect that this estimate can be sharp in the sense that there may exist a
flow field with arbitrary P\'eclet number that may realize it.  Certainly this
is true in $2d$: given the source-sink function $s(\bm{x})$ with unstirred
scalar distribution $\theta_{0}(\bm{x})$, just define a flow field with stream
function $\psi(\bm{x}) \sim \theta_{0}(\bm{x})$.  The streamlines of such a
flow are along level sets of $\theta_{0}$ so the flow has no effect.  Indeed,
for this flow $\bm{u}\cdot \nabla \theta_{0} = 0$ no matter what the magnitude
of $U$ is, so $\theta_{0}$ is the stationary solution of the
advection--diffusion equation for any value of $\Pe$.  However, this perfectly
``non-mixing'' flow is not statistically stationary, homogeneous and
isotropic, and it is not clear whether further constraints derived from the
full advection--diffusion equation might be implemented to raise this lower
estimate for such fluctuating flows.  We leave that question for a future
study.

We can follow the same line of reasoning to derive lower estimates on the
other mixing efficiencies, although perhaps with less satisfaction.  A lower
bound on ${\cal E}_{0}$ requires an upper estimate on $\langle \theta^{2}
\rangle$.  Starting from (\ref{elphaba2}), recalling that $\theta$ has spatial
mean zero and invoking Poincar\'e's inequality on the left and Cauchy--Schwarz
on the right,
\begin{equation}
\kappa \frac{2\pi}{L} \langle \theta^{2} \rangle^{1/2}
\langle |\nabla \theta|^{2} \rangle^{1/2} \le
\kappa \langle |\nabla \theta|^{2} \rangle = \langle s \theta \rangle
\le \langle |\nabla^{-1}s|^{2} \rangle^{1/2} \langle |\nabla \theta|^{2}
\rangle^{1/2}.
\label{elphaba3}
\end{equation}
Hence we deduce the upper estimate on the scalar variance,
\begin{equation}
\langle \theta^{2} \rangle \le
\frac{L^{2}}{4 \pi^{2} \kappa^{2}} \langle |\nabla^{-1}s|^{2} \rangle.
\label{elphaba4}
\end{equation}
The unstirred variance is
\begin{equation}
\langle \theta_{0}^{2} \rangle =
\frac{1}{\kappa^{2}} \langle (\Delta^{-1} s)^{2} \rangle,
\label{elphaba5}
\end{equation}
so we have the lower estimate
\begin{equation}
{\cal E}_{0}^{2} =
\frac{\langle \theta_{0}^{2} \rangle}{\langle \theta^{2} \rangle} \ge
\frac{4 \pi^{2}}{L^{2}}
\frac{\langle (\Delta^{-1} s)^{2} \rangle}{\langle |\nabla^{-1}s|^{2} \rangle}
= \frac{\sum_{{\bm{k}}} (Lk/2\pi)^{-4}|\hat{s}_{\bm{k}}|^{2}}
{\sum_{{\bm{k}}} (Lk/2\pi)^{-2}|\hat{s}_{\bm{k}}|^{2}}.
\label{elphaba6}
\end{equation}
This lower estimate is strictly positive but because $Lk/2\pi \ge 1$ it is
$\le 1$.  Hence it does not rule out the existence of flows that might {\it
increase} scalar variance.  Note as well that this estimate depends
explicitly on the functional ``shape'' of the source-sink distribution, a
feature that we will find for many estimates---upper and lower---on the mixing
efficiencies.

The result can in principle be sharpened by solving the variational problem
\begin{equation}
\langle \theta^{2} \rangle \le \max_{\vartheta} \,
\{ \langle \vartheta^{2} \rangle \, | \,
\kappa \langle |\nabla \vartheta|^{2} \rangle = 
\langle s \vartheta \rangle
\}
\label{elphaba7}
\end{equation}
where the maximization is performed over all $\vartheta(\bm{x})$ satisfying
the (periodic) boundary conditions on the domain.
The Euler--Lagrange equation for the maximizer $\vartheta_{*}(\bm{x})$ is
\begin{equation}
0 = 2 \vartheta_{*} - 2\mu \kappa \Delta \vartheta_{*} - \mu s(\bm{x})
\label{elphaba8}
\end{equation}
where $\mu$ is the Lagrange multiplier enforcing the constraint
(\ref{elphaba2}).  In terms of the Fourier coefficients the solution of
(\ref{elphaba8}) is straightforward,
\begin{equation}
\hat{\vartheta_{*}}_{\bm{k}} = \frac{\mu}{2} \, \frac{\hat{s}_{\bm{k}}}
{\mu \kappa k^{2} + 1},
\label{elphaba9}
\end{equation}
but $\mu$ is the solution of 
\begin{equation}
\frac{1}{2} \sum_{{\bm{k}}} \frac{\mu \kappa k^{2}
|\hat{s}_{\bm{k}}|^{2}}{(\mu \kappa k^{2}+ 1)^{2}} =
\sum_{{\bm{k}}} \frac{|\hat{s}_{\bm{k}}|^{2}}{\mu \kappa k^{2}+ 1}.
\label{elphaba10}
\end{equation}
In general it is difficult to solve (\ref{elphaba10}) for $\mu$,
but there is one case where it is easy: if the source is 
``monochromatic'', i.e., involves only a single
wavenumber of amplitude $k_{s}$, then $\mu = -2/\kappa k_{s}^{2}$ so
\begin{equation}
{\cal E}_{0}^{2} \ge \frac{\sum_{{\bm{k}}} |\hat{\theta}_{0\bm{k}}|^{2}}
{\sum_{{\bm{k}}} |\hat{\vartheta_{*}}_{\bm{k}}|^{2}}
= \frac{4\sum_{{\bm{k}}} |\hat{s}_{\bm{k}}|^{2}/(\mu \kappa k^{2})^{2}}
{\sum_{{\bm{k}}} |\hat{s}_{\bm{k}}|^{2}/(\mu \kappa k^{2}+ 1)^2}
= 1.
\label{elphaba11}
\end{equation}
Hence stirring a monochromatic source can never increase the variance.
However if the source-sink distribution involves even just two distinct
wavenumbers, then the solution of (\ref{elphaba10})---which must be performed
numerically---yields a value for $\mu$ that produces a lower bound for
${\cal E}_{0}$ that is strictly less than 1~\cite{ShawGFD2005}.  
Further details are relegated to Appendix A; this analysis does not 
prove that there actually is some stirring that can
increase the scalar variance for such sources, but it leaves open the
possibility.  For the purposes of this study we will settle for this lower
bound as far as its P\'eclet number scaling is concerned.  That is,
\begin{equation}
{\cal E}_{0} \ge c \, \Pe^{0},
\label{elphaba12}
\end{equation}
where the positive coefficient $c$ may depend on the source-sink
distribution, and we cannot rule out that it might be less than 1.

A lower bound on the large-scale mixing efficiency follows from the constraint
(\ref{elphaba2}) via Poincar\'e's and the Cauchy--Schwarz inequalities as well.
It follows from (\ref{elphaba4}) that
\begin{equation}
\frac{16\pi^{4}}{L^{4}} \langle |\nabla^{-1}\theta|^{2} \rangle \le
\frac {\langle |\nabla^{-1} s|^{2} \rangle} {\kappa^{2}}
\label{elphaba13}
\end{equation}
so that
\begin{equation}
{\cal E}_{-1}^{2} \ = \
\frac{\langle |\nabla^{-1}\theta_{0}|^{2} \rangle}{\langle |\nabla^{-1}\theta|^{2} \rangle}
\ge \frac{16 \pi^{4}}{L^{4}} \frac{\langle |\nabla^{-3} s|^{2} \rangle}{ \langle |\nabla^{-1} s|^{2} \rangle}
\ = \ \frac{\sum_{{\bm{k}}} (Lk/2\pi)^{-6}|\hat{s}_{\bm{k}}|^{2}}
{\sum_{{\bm{k}}} (Lk/2\pi)^{-2}|\hat{s}_{\bm{k}}|^{2}} \ \le \ 1.
\label{elphaba14}
\end{equation}
We note again that for the special case of a monochromatic source, a
variational formulation as in (\ref{elphaba7}) yields the improved lower bound
${\cal E}_{-1} \ge 1$.
It remains an open problem to determine if these lower estimates $\sim \Pe^{0}$
are sharp for arbitrary sources and sinks stirred by some statistically homogeneous 
and isotropic flow.  (We will show that they are sharp in some particular cases.)
Certainly it will be necessary to use more than just (\ref{elphaba2}) to answer
this question in general.

\subsection{Upper bounds on ${\cal E}_{0}$}
\label{sec:bound0}

This analysis begins by multiplying the advection--diffusion equation
\eqref{eq:AD} by a smooth, time-independent, spatially periodic ``projector
function'' $\varphi(\bm{x})$ and taking the space-time average and integrating
by parts to obtain
\begin{equation}
  0 \ = \ \langle \theta (\bm{u}\cdot \nabla + \kappa \lapl)\varphi \rangle
  +\langle \varphi s \rangle.
  \label{eq:firststep}
\end{equation}
Because this constraint holds for all~$\varphi$, a lower bound on the variance
is
\begin{equation}
  \langle \theta^2 \rangle \ \geq \ \max_{\varphi} \min_{\vartheta}
  \{ \langle \vartheta^2 \rangle~|~0 = \langle \vartheta
   (\bm{u} \cdot \nabla \varphi + \kappa \lapl \varphi) \rangle +
   \langle \varphi s \rangle \}
\label{lbvar}
\end{equation}
where $\vartheta(\bm{x},t)$ varies over all spatially periodic function with
unconstrained dependence.\footnote{Previously, Thiffeault, Doering
\& Gibbon~\cite{Thiffeault2004} derived a
bound on this variance without optimizing over $\varphi$.}  
This min-max variational formulation is equivalent to
 \begin{equation}
  \langle \theta^2 \rangle \ \geq \ \min_{\vartheta }
  \{ \langle \vartheta^2 \rangle~|~ 
  \overline{\bm{u} \cdot \nabla \vartheta } = 
  \kappa \lapl \overline{\vartheta} + s\}.
 \label{glinda1}
\end{equation}
The multiplier function $\varphi(\bm{x})$ plays the role of a Lagrange
multiplier to impose the ``Reynolds averaged'' advection--diffusion equation
$\overline{\bm{u} \cdot \nabla \vartheta } = \kappa \lapl \overline{\vartheta}
+ s$ as a constraint.  The formulation of the bound as a min-max 
problem in (\ref{lbvar}) is, as we will see, convenient for its solution.

The minimization over $\vartheta$ in (\ref{lbvar}) is equivalent to an
application of the Cauchy--Schwarz inequality to \eqref{eq:firststep}:
\begin{equation}
  \langle \theta^2 \rangle \geq \max_{\varphi} \frac{ \langle \varphi s
  \rangle^2}{\langle (\bm{u}\cdot \nabla \varphi+ \kappa \lapl\varphi)^2
  \rangle} = \max_{\varphi} \frac{ \langle \varphi s \rangle^2}{\langle
  \varphi\, \overline{{\cal{L}}{\cal{L}}^\dagger} \varphi \rangle}\,,
  \label{eq:varb}
\end{equation}
where we defined the advection--diffusion operator~$\cal{L}$ and its
adjoint~${\cal{L}}^\dagger$,
\begin{equation}
  {\cal{L}}\ldef \bm{u}\cdot \nabla -\kappa
  \lapl \qquad \text{and} \qquad
  {\cal{L}}^\dagger\ldef -\bm{u}\cdot \nabla -\kappa
  \lapl\,.
\end{equation}
We explicitly indicate the time average in the denominator
of~\eqref{eq:varb} remembering that~$\varphi$ is time-independent.  
An important point here is that we can average the time-dependent
self-adjoint operator ${\cal{L}}{\cal{L}}^\dagger$ before
carrying out the maximization over~$\varphi$.

Maximizing~\eqref{eq:varb} over $\varphi$ is equivalent to minimizing its
denominator.  Without loss of generality, since the functional~\eqref{eq:varb}
is homogeneous in~$\varphi$, we constrain $\varphi$ to have unit projection
onto the source.  Thus we must minimize the functional
\begin{equation}
  {\cal{F}}\ldef\left \langle \tfrac{1}{2} \varphi\,
  \overline{{\cal{L}}{\cal{L}}^\dagger} \varphi -\mu (\varphi s -1) \right
  \rangle,
\end{equation}
leading to the Euler--Lagrange equation
\begin{equation}
  0 = \frac{\delta{\cal{F}}}{\delta \varphi}=
  \overline{{\cal{L}}{\cal{L}}^\dagger}\, \varphi - \mu s
\end{equation}
where $\mu$ is a Lagrange multiplier to enforce the constraint $\langle \varphi s \rangle = 1$.  
The minimizer is then
\begin{equation}
  \varphi = \frac{(\overline{{\cal{L}}{\cal{L}}^\dagger})^{-1} s}
  {\langle s (\overline{{\cal{L}}{\cal{L}}^\dagger})^{-1} s \rangle}.
  \label{eq:phiopt0}
\end{equation}
Inserting~\eqref{eq:phiopt0} into~\eqref{eq:varb}, we obtain the lower bound
\begin{equation}
  \langle \theta^2 \rangle \geq \langle s
  (\overline{{\cal{L}}{\cal{L}}^\dagger})^{-1} s \rangle =
  \langle s \,\{ \kappa^2 \lapl^2 - \nabla
  \cdot(\overline{\bm{u}\bm{u} })+\kappa(2 \nabla
  \overline{\bm{u}}:\nabla \nabla + \lapl \overline{\bm{u}} \cdot
  \nabla) \}^{-1} s\rangle.
  \label{eq:sM0s}
\end{equation}
Interestingly, this estimate depends only on the mean and equal-point
correlation of the stirring.

Specializing to flows satisfying the assumptions of statistical homogeneity
and isotropy in (\ref{mean1}) and (\ref{eq:SHIT})---actually we just use
(\ref{mean1}) and (\ref{eq:HIT1}) here---we can carry out the time average in
\eqref{eq:sM0s}, yielding
\begin{equation}
  \langle \theta^2 \rangle \geq
  \langle s(\overline{{\cal{L}}{\cal{L}}^\dagger})^{-1}s \rangle
  = \langle s\, \{ \kappa^2 \lapl^2 -
  (U^2/d)\lapl \}^{-1} s\rangle
  = \sum_{\bm{k}} \frac{|\hat{s}_{\bm{k}}|^2}{\kappa^{2} k^4 +U^2 k^2/d}.
  \label{eq:var0HITbound}
\end{equation}
Using $\theta_{0} = (-\kappa \Delta)^{-1}s$ we express this result as an upper
bound on the mixing efficiency ${\cal E}_{0}$:
\begin{equation}
{\cal E}_{0}^2 = \frac{\langle \theta_{0}^{2} \rangle}{\langle \theta^{2} \rangle}
\leq \frac{\langle s\, \lapl^{-2}\,s \rangle} 
{\langle s\, (\lapl^2 - (\Pe^2/L^2 d)\,\lapl)^{-1}\,s \rangle}
= \frac{\sum_{\bm{k}} |\hat{s}_{\bm{k}}|^2/k^4}
{\sum_{\bm{k}} |\hat{s}_{\bm{k}}|^2/(k^4 +k^2 \Pe^2 / L^{2}d)}
\label{eq:varlubound}
\end{equation}
where the dimensionless P\'eclet number $\Pe = UL/\kappa$ has been inserted.
We observe that like the lower estimate on ${\cal E}_{0}$ in (\ref{elphaba6}),
the upper bound on the mixing efficiency depends on the shape of the
source-sink distribution.

\subsection{Upper bounds on ${\cal E}_{1}$}
\label{sec:bound1}

The gradient variance $\langle |\nabla \theta|^2 \rangle$ can quickly and easily
be bounded from below in a similar manner to the variance.
Begin with Eq.~\eqref{eq:firststep}, integrate by parts, and apply the
Cauchy--Schwarz inequality to obtain
\begin{equation}
  \langle \varphi s \rangle^2 = \langle ( \bm{u} \varphi + \kappa \nabla
  \varphi) \cdot \nabla \theta \rangle^2 \leq \langle | \bm{u} \varphi +
  \kappa \nabla \varphi|^2 \rangle \langle | \nabla \theta|^2 \rangle
  \label{eq:constr1}
\end{equation}
so that
\begin{equation}
  \langle |\nabla \theta |^2 \rangle \geq \max_{\varphi} \frac{\langle \varphi
  s \rangle^2}{\langle |\bm{u} \varphi + \kappa \nabla \varphi|^2
  \rangle}.
  \label{eq:gvbound}
\end{equation}
The right-hand side of~\eqref{eq:gvbound} is homogeneous in~$\varphi$
so we minimize the denominator subject to the constraint that 
$\langle \varphi s \rangle = 1$.  
Under the homogeneity and isotropy assumptions  (\ref{mean1}) and
(\ref{eq:HIT1}), the challenge becomes to evaluate
\begin{equation}
  \min_{\varphi} ~\{\langle \kappa | \nabla \varphi|^2 + U^2\varphi^2 \rangle
  ~|~ \langle \varphi s \rangle=1\}.
\end{equation}
Following a similar development as in Section~\ref{sec:bound0}, the solution is found to be
\begin{equation}
\langle |\nabla \theta |^2 \rangle \geq
\langle s (-\kappa^2 \lapl  + U^2)^{-1} s \rangle
\end{equation}
and the mixing efficiency at small scales is bounded according to
\begin{equation}
  {\cal E}_{1}^{2} \leq \frac{ \langle s\,(-\lapl)^{-1}\,s
  \rangle}{\langle s (-\lapl  + \Pe^2/L^2)^{-1} s \rangle}
  = \frac{\sum_{\bm{k}} |\hat{s}_{\bm{k}}|^2/k^2}
{\sum_{\bm{k}} |\hat{s}_{\bm{k}}|^2/(k^2 + \Pe^2 / L^{2})}.
\label{E0cheap}
\end{equation}
Like the upper bound for ${\cal E}_{0}$ in \eqref{eq:varlubound}, this
estimate depends on the functional structure of the sources and sinks, and on
the statistically homogeneous and isotropic stirring only through the P\'eclet
number.

We can improve this upper estimate by avoiding the application of the
Cauchy--Schwartz inequality that led to~\eqref{eq:constr1}.  We used that
inequality above for expediency, but in fact we expect the bound to involve
only the gradient (i.e., curl-free) part of the field $\bm{u}\varphi+\kappa
\nabla \varphi$.  This can be seen by solving the full min-max variational
problem
\begin{equation}
  \langle |\nabla \theta |^2 \rangle \geq
  \max_{\varphi} \min_{\vartheta}
  \{ \langle | \nabla \vartheta|^2 \rangle ~|~
  \langle \varphi s \rangle = \langle (\bm{u} \varphi + \kappa \nabla
  \varphi) \cdot \nabla \vartheta \rangle \}.
\label{337}
\end{equation}
The minimization is straightforward:
\begin{equation}
  \langle | \nabla \theta |^2 \rangle  \ge \max_{\varphi}
  \frac{ \langle \varphi s \rangle^2}{\langle (\nabla \cdot \bm{w})
  (-\lapl^{-1})\nabla \cdot \bm{w} \rangle}
  \label{eq:varsolsharper}
\end{equation}
where $\bm{w}=\bm{u}\varphi + \kappa \nabla \varphi$.
The vector field $\bm{w}$ can be decomposed as
\begin{eqnarray}
  &&\bm{w}=\underbrace{\bm{w} -\nabla \lapl^{-1} \nabla \cdot
  \bm{w} }+ \underbrace{\nabla \lapl^{-1} \nabla \cdot \bm{w}}\\
  &&~~~~~~~{{\text{divergence-free}}~~~~~~~{\text{curl-free}}}\nonumber
\end{eqnarray}
where the two components are orthogonal.
Only the curl-free part contributes in (\ref{eq:varsolsharper}).
That is, the denominator is just the norm of the curl-free portion:
\begin{align}
  \langle (\nabla \cdot \bm{w}) (-\lapl^{-1}) (\nabla \cdot \bm{w})
  \rangle &= \langle [\lapl \lapl^{-1} (\nabla \cdot
  \bm{w})](-\lapl^{-1}) (\nabla \cdot \bm{w}) \rangle \nonumber\\
  &= \langle \nabla (\lapl^{-1} (\nabla
  \cdot \bm{w})) \cdot \nabla (\lapl^{-1} (\nabla \cdot \bm{w}))\rangle
  \nonumber\\
  &= \langle | \nabla \lapl^{-1} \nabla \cdot \bm{w}|^2 \rangle
  \ \le \ \langle |\bm{w}|^2 \rangle.
\end{align}
Hence (\ref{eq:varsolsharper}) generally represents an improvement over the
expression resulting from application of the Cauchy--Schwarz inequality in
(\ref{eq:gvbound}).

This improved bound depends on the full two-point correlation function of the
velocity field:
\begin{eqnarray}
  \langle (\nabla \cdot \bm{w}) (-\lapl^{-1})\nabla \cdot \bm{w}
  \rangle &=& \frac{1}{L^d} \int \dint^{d}x \int \dint^{d}y \
  \overline {\nabla_{\bm{x}} \cdot \bm{w}(\bm{x},\cdot) 
  G( \bm{x} - \bm{y}) \nabla_{\bm{y}} \cdot
  \bm{w}(\bm{y},\cdot)} \nonumber \\
  &=&\frac{1}{L^d} \int \dint^{d}x \int \dint^{d}y \
  (-\nabla_{\bm{x}} \nabla_{\bm{x}} G):
  \overline{\bm{w}(\bm{x},\cdot)\bm{w}(\bm{y},\cdot)}
\end{eqnarray}
where $G(\bm{x}-\bm{y})$ is the Green's function for $-\lapl$ on spatially
mean-zero functions with Fourier coefficients $\hat{G}(k)=1/(L^{d}k^{2})$ for
$\bm{k} \ne 0$.
%Integrating by parts,
%\begin{equation}
%  \langle (\nabla \cdot \bm{w}) (-\lapl^{-1})\nabla \cdot \bm{w}
%  \rangle = \frac{1}{L^d} \int \dint^{d}x \int \dint^{d}y \
%  (-\nabla_{\bm{x}} \nabla_{\bm{x}} G):
%  \overline{\bm{w}(\bm{x},\cdot)\bm{w}(\bm{y},\cdot)}.
%\end{equation}
Under the assumptions (\ref{mean1}) and (\ref{eq:SHIT}) 
of statistical homogeneity and isotropy,
\begin{eqnarray}
  \overline{w_{i}(\bm{x},\cdot)w_{j}(\bm{y},\cdot)} &=&
  \varphi(\bm{x}) \varphi(\bm{y})
  \overline{u_{i}(\bm{x},\cdot)u_{j}(\bm{y},\cdot)} +
  \kappa^2 \partial_{i}\varphi(\bm{x}) \partial_{j}\varphi(\bm{y})
  \nonumber \\
  &=& C_{ij}(\bm{x}-\bm{y})\varphi(\bm{x}) \varphi(\bm{y})+
  \kappa^2 \partial_{i}\varphi(\bm{x}) \partial_{j}\varphi(\bm{y}).
\end{eqnarray}
In terms of Fourier transformed variables,
\begin{equation}
  \langle (\nabla \cdot \bm{w}) (-\lapl^{-1})\nabla \cdot \bm{w} \rangle = 
  \sum_{\bm{k},\bm{k'} \ne 0} \frac{\hat{C}(k)}{d-1}
  \left( 1-\frac{(\bm{k} \cdot \bm{k'})^{2}}{k^{2}k'^{2}}\right)
  | \hat{\varphi}_{\bm{k}+\bm{k'}}|^{2} +
  \sum_{\bm{k} \ne 0} \left( \frac{\hat{C}(0)}{d} + \kappa^{2} k^{2} \right)
  | \hat{\varphi}_{\bm{k}}|^{2}.
\end{equation}
Thus the mixing efficiency at small scales is actually bounded from above according to
\begin{eqnarray}
{\cal E}_{1}^{2} &\leq&
\min_{\varphi} \ \frac{\sum_{\bm{k}} |\hat{s}_{\bm{k}}|^2/k^2}
{\left[ \sum_{\bm{k}} \hat{s}_{\bm{k}}^{*}\hat{\varphi}_{\bm{k}}\right]^{2} }
\ \times \nonumber \\
&\times& 
\left[
 \sum_{\bm{k},\bm{k'} \ne 0} \frac{\hat{C}(k)}{(d-1) \kappa^{2} }
  \left( 1-\frac{(\bm{k} \cdot \bm{k'})^{2}}{k^{2}k'^{2}}\right)
  | \hat{\varphi}_{\bm{k}+\bm{k'}}|^{2} +
  \sum_{\bm{k} \ne 0} \left( \frac{\hat{C}(0)}{d \kappa^{2}} + k^{2} \right)
  | \hat{\varphi}_{\bm{k}}|^{2}.
\right]
\label{IB}
\end{eqnarray}
This upper bound depends on details of the full spectrum of the stirring
velocity field.

We will not perform the optimization over $\varphi$ for the general problem
here; the implications of two-point statistical properties of the stirring on
the P\'eclet number dependence of this bound on ${\cal E}_{1}$ will be left
for future investigations.  There is one case, however, where the optimization
can easily be carried out that shows that the result of (\ref{IB}) may indeed
be a quantitative improvement over (\ref{E0cheap}).  That simple case is when
the spectrum of the velocity field is concentrated at $\bm{k} = 0$, i.e., when
the velocity field is at (almost) every moment of time a uniform ``wind'' in
space.  For such spatially uniform statistically homogeneous and isotropic
flows, $\hat{C}(0) = U^{2}$ while all the other $\hat{C}(k) = 0$ for $\bm{k}
\ne 0$.  Then
\begin{eqnarray}
{\cal E}_{1}^{2} \ \leq \
\min_{\varphi} \ \frac{\sum_{\bm{k}} |\hat{s}_{\bm{k}}|^2/k^2}
{\left[ \sum_{\bm{k}} \hat{s}_{\bm{k}}^{*}\hat{\varphi}_{\bm{k}} \right]^{2}}
  \sum_{\bm{k} \ne 0} \left( \frac{U^{2}}{d \kappa^{2}} + k^{2} \right)
  | \hat{\varphi}_{\bm{k}}|^{2}
 \ = \ \frac{\sum_{\bm{k}} |\hat{s}_{\bm{k}}|^2/k^2}
{\sum_{\bm{k}} |\hat{s}_{\bm{k}}|^2/(k^2 + \Pe^2 / dL^{2})}.
\label{IB2}
\end{eqnarray}
In this case the improvement over (\ref{E0cheap}) is just the extra factor of
the spatial dimension $d$ in the denominator of the denominator of the
denominator in the last term.

\subsection{Upper bounds on ${\cal E}_{-1}$}
\label{sec:boundm1}

To derive a lower bound on the inverse-gradient variance
$\langle |\nabla^{-1} \theta|^2 \rangle$ we begin
again with (\ref{eq:firststep}), insert $\lapl\lapl^{-1}=1$, 
integrate by parts and apply the Cauchy--Schwarz
inequality to obtain
\begin{align}
  \langle \varphi s \rangle \ = \ \langle \nabla (\bm{u} \cdot \nabla \varphi +
  \kappa \lapl \varphi) \cdot \nabla \lapl^{-1} \theta \rangle
  % \nonumber \\
  \ \leq \ \langle | \nabla (\bm{u} \cdot \nabla \varphi + \kappa \lapl
  \varphi) |^2 \rangle ^{\frac{1}{2}} \langle | \nabla^{-1} \theta| ^2 \rangle
  ^{\frac{1}{2}}.
\end{align}
This gives a lower bound on the inverse-gradient variance
\begin{equation}
  \langle | \nabla^{-1} \theta|^2 \rangle \geq\max_{\varphi} \frac{ \langle
  \varphi s \rangle^2 }{ \langle | \nabla(\bm{u}\cdot \nabla
  \varphi+\kappa \lapl \varphi|^2\rangle}\,.
  \label{Oz}
\end{equation}
Recalling that $\varphi$ is time-independent and restricting attention to statistically
homogeneous and isotropic flows---assuming as well that the enstrophy 
$\Omega^{2} = \langle |\nabla \bm{u}|^{2} \rangle$ is finite---the
denominator is
\begin{equation}
  |\nabla \bm{u} \cdot \nabla \varphi+ \bm{u} \cdot \nabla \nabla
  \varphi + \kappa \nabla \lapl \varphi  |^2 =(\Omega^2/d) | \nabla \varphi|^2 + (U^2/d)(\lapl
  \varphi)^2+ \kappa^2
  |\lapl \nabla \varphi|^2 .
\end{equation}
Then the optimization over $\varphi$ in (\ref{Oz}) is straightforward.  Recalling that
\begin{equation}
\langle | \nabla^{-1} \theta_0|^2 \rangle =
\langle  | \nabla^{-1} \lapl^{-1} s |^2 \rangle/\kappa^{2}
\end{equation}
and the definition $\lambda = U/\Omega$,
we conclude that
\begin{eqnarray}
{\cal E}_{-1}^2 &\leq& \frac{ \langle
  | \nabla^{-1} \lapl^{-1} s |^2 \rangle}
       { \langle s \l(-\lapl^3 + (\Pe^2/L^2\,d)
\lapl^2 - (\Pe^2/\lambda^2 L^2\,d) \lapl\r)^{-1} s \rangle}
\nonumber \\
&=& 
\frac{\sum_{\bm{k}} |\hat{s}_{\bm{k}}|^2/k^6}
{\sum_{\bm{k}} |\hat{s}_{\bm{k}}|^2/(k^6 +k^4 \Pe^2 / L^{2}d
+k^2 \Pe^2 / \lambda^{2} L^{2}d)} .
\label{eq:igvarlubound}
\end{eqnarray}

Compare this to (\ref{eq:varlubound}) for the upper estimate on ${\cal E}_{0}$
and (\ref{E0cheap}) and (\ref{IB}) for the upper estimate on ${\cal E}_{1}$.
Note that the efficiency depends on the enstropy in the flow, i.e., the flow's
``shear'' or ``strain'' content, directly through $\lambda = U/\Omega$.
Interestingly, the $\Omega$ (or $\lambda$) term allows for an increase in the
bound on the mixing efficiency on large scales via stirring on small scales,
something that is absent in the bounds on the mixing efficiency at
intermediate and small scales.

%===================================================

\section{Saturating the Multi-scale Mixing Efficiency Bounds}
\label{sec:saturation}

The upper bounds on the mixing efficiencies ${\cal E}_{p}$ derived in the
previous section depend on the entire source-sink distribution functions, but
on just a few features ($U$, and for ${\cal E}_{-1}$, $\lambda=U/\Omega$) of
the statistically homogeneous and isotropic stirring field.  In this section
we show that there is at least one combination of sources, sinks and stirring
strategies that saturate the upper estimates exactly at all P\'eclet numbers.
This establishes that on the highest level of generality the upper bound
analysis is absolutely sharp.  We also show that there is at least one
combination of sources, sinks and statistically stationary homogeneous
isotropic stirrings that saturate the scaling (if not necessarily the
prefactor) of the lower estimates on the efficiencies $\sim \Pe^{0}$.

Consider the simple monochromatic source function
\begin{equation}
s(\bm{x}) \ = \ \sqrt{2} S \sin(k_{s}x_{1})
\label{MS}
\end{equation}
where $S$ is the root mean square amplitude, $2\pi/k_{s}$ is the wavelength,
and $x_{1}$ is one of the $d$ coordinates.
For this monochromatic source the lower bounds on the mixing efficiencies
at all scales are the same, i.e., for $p = -1, 0, 1$,
\begin{equation}
1 \ \le \ {\cal E}_{p} \quad
\text{ (monochromatic sources)}.
\end{equation}

The upper bounds on ${\cal E}_{0}$, ${\cal E}_{1}$ and ${\cal E}_{-1}$ in,
respectively, (\ref{eq:varlubound}), (\ref{E0cheap}) and (\ref{IB2}) are
generally different for monochromatic source-sink distribution functions.  But
if we restrict attention to ``uniform wind'' flows $\bm{u}(\bm{x},t)$ that are
at each instant of time spatially uniform (i.e., $\nabla \bm{u}(\bm{x},t)=0$),
then $\Omega = 0$, $\lambda \rightarrow \infty$, and we can take advantage of
the improved bound for ${\cal E}_{1}$ in (\ref{IB2}) to see that the upper
estimates are all the same.  For $p = -1, 0, 1$,
\begin{equation}
{\cal E}_{p} \ \leq \ \sqrt{1+\Pe^2/ L^{2}k_{s}^{2}d}
\quad \text{ (monochromatic sources \& uniform winds)}.
\label{UMU}
\end{equation}

In order to show that the upper bounds in (\ref{UMU}) are sharp, we construct
a family of statistically homogeneous and isotropic flows that approach these
limits.  The trick is to sustain a uniform wind in a given direction for a
``long'' time so that the steady state is very nearly achieved before
switching to a uniform wind in another direction.  The dynamic transients
between the changes in flow configurations decay at least at rate $\kappa
k_{s}^{2}$, so when the transitions are sufficiently infrequent the scalar
field is almost always (nearly) in a static configuration.  Then we may solve
the steady flow problem for the scalar field exactly and average the variances
over the wind directions to evaluate the efficiencies in the limit of slow
switches among the directions.  This adiabatic averaging method can be
implemented for other flows, too, as will be considered in
Section~\ref{sec:SSSSSF}.

Consider uniform winds of speed $U$ blowing along the ``diagonals'', in the
$2^{d}$ directions given by unit vectors $\frac{1}{\sqrt{d}} \left( \pm
\hat{\bm{e}}_{1} \dots \pm \hat{\bm{e}}_{d} \right)$.  Each of these
directions is equivalent, so we can solve any single problem to evaluate the
variances.  The steady advection--diffusion equation with source
$s(\bm{x})=\sqrt{2}S \sin (\ks x_1)$ and uniform stirring field
$\bm{u}(\bm{x})=\frac{U}{\sqrt{d}}\sum_{j=1}^{d} \hat{\bm{e}}_{j}$ is
\begin{equation}
  \frac{U}{\sqrt{d}} \sum_{j=1}^d \frac{\partial \theta}{\partial x_j} =
  \kappa \sum_{j=1}^d \frac{\partial^2 \theta}{\partial x_j^2} + \sqrt{2} S
  \sin(\ks x_1).
\end{equation}
The solution is of the form $\theta(\bm{x})=\sum_{j=1}^dF^{(j)}(x_j)$ where
the functions $F^{(j)}(x_j)$ satisfy the system of constant coefficient ODEs
\begin{equation}
\begin{split}
  \frac{d^2F^{(1)}}{dx_1^2} - \frac{U}{\sqrt{d}\,\kappa}\, \frac{dF^{(1)}}{dx_1}
  +\frac{\sqrt{2}S}{\kappa}\, \sin(\ks x_1/L)&=0,\\
  \frac{d^2F^{(j)}}{dx_n^2} - \frac{U}{\sqrt{d}\,\kappa}\, \frac{dF^{(j)}}{dx_n}
  &=0,\qquad\text{for $2 \leq j \leq d$,}
\end{split}
\end{equation}
with periodic boundary conditions $F^{(j)}(0)=F^{(j)}(L)$.  Recalling that
$\theta$ has, without loss of generality, spatial mean zero, the solution is
\begin{equation}
\begin{split}
  F^{(1)} &= \frac{\sqrt{2}S}{\kappa k_{s}^{2} + U^2/d\kappa}\left[
    \sin(\ks x_1) - \frac{U}{\kappa k_{s} \sqrt{d}}\, \cos(\ks x_1) \right],\\
  F^{(j)} &= 0,\qquad\text{for\ } 2\leq j \leq d.
\end{split}
\end{equation}
The variance is thus
\begin{equation}
  \langle \theta^2 \rangle = \frac{S^2}{\kappa^{2} k_{s}^{4}
  + k_{s}^{2}U^{2}/d},
\end{equation}
and because the scalar field is monochromatic, the small-scale and large-scale
variances simply satisfy $k_{s}^{2} \langle |\nabla^{-1} \theta|^2 \rangle =
\langle |\nabla \theta|^2 \rangle/k_{s}^{2} =\langle \theta^2 \rangle$.  The
multi-scale mixing efficiencies are all then
\begin{equation}
{\cal E}_{p} = \sqrt{\frac{\langle  |\nabla^{p} \theta_{0}|^2 \rangle}
{\langle  |\nabla^{p} \theta|^2 \rangle}}
= \sqrt{1+ \frac{\Pe^2}{k_{s}^{2}L^{2}d}}
\end{equation}
precisely as in (\ref{UMU}).

On one hand this result is fairly intuitive: the most efficient way to reduce the variance 
(on any length scale) is to direct the flow from source regions directly toward the 
closest convenient sink regions and the likewise from sinks toward sources---{\it if} 
this can be accomplished effectively given the constraints of incompressibility 
and statistical homogeneity and isotropy.  
This can be done simply for these monochromatic flows on the torus, 
and we have discovered that such flows actually suppress the scalar
variance at all scales as well as is possible for {\it any} 
statistically homogeneous and isotropic flow field. 
(Note: this example was inspired by Plasting \& Young~\cite{Plasting2006}
who showed that the steady direct flow across the source saturates the variance 
bound for single-wavenumber sources among {\it all} flows regardless
of any statistical considerations.)
On the other hand this type of sweeping flow is somewhat pathological in the 
sense that it simply transports the source onto the sink without ``stirring''
and ``mixing'' by the usual meanings of those words.
In some cases, such as transient mixing as discussed in the introduction, the 
role of effective stirring is to stretch material lines and amplify gradients 
of the passive scalar to accelerate the action of molecular diffusion 
to dissipate variance on small scales.

It is not presently clear whether the upper bounds in (\ref{eq:varlubound}),
(\ref{IB}) and (\ref{IB2}) can be saturated for more general
sources and sinks.
Moreover, this particular saturation result depends explicitly on the 
geometry of the domain.
In geophysical and astrophysical applications, for example, it is natural
to consider the domain to be the surface of a sphere.  
While such a domain admits similar concepts of statistical homogeneity
and isotropy for the flow and ``monochromaticity'' for the source-sink 
distribution (eigenfunctions of the Laplacian on the sphere), there is
no analogous uniform sweeping flow as there is on the torus.  
This makes it clear that the optimal stirrer is a function of both
the source shape and the domain.
Formulating the optimization problem for the best stirring field for a given
source-sink distribution remains a problem for future investigation.

We close this section by pointing out that a similar statistically homogeneous
and isotropic uniform wind can also be arranged to saturate the high-$\Pe$ scaling
of the {\em lower} bounds on the mixing efficiencies.  For the monchromatic source
(\ref{MS}), consider uniform winds oriented along the coordinate axes in the
$2d$ directions $\pm \hat{\bm{e}}_{j}$, switching only after blowing each way
for a long enough time for the transients to be negligible for the time
averages.  The wind reduces the scalar variances below the variance in
$\theta_{0}$ only when it blows in the $\pm \hat{\bm{e}}_{1}$ directions,
which occurs $1/d$ of the time.  Averaging the scalar variance adiabatically
over the wind directions yields the mixing efficiencies
\begin{equation}
{\cal E}_{p} =
\sqrt{\frac{1+\Pe^{2}/k_{s}^{2}L^{2}}{1+(d-1)\Pe^{2}/dk_{s}^{2}L^{2}}}.
\end{equation}
These efficiencies are monotonically increasing in $\Pe$ but bounded according
to
\begin{equation}
1 \ \le \ {\cal E}_{p} \ < \ \lim_{\Pe \rightarrow \infty}
\sqrt{\frac{1+\Pe^{2}/k_{s}^{2}L^{2}}{1+(d-1)\Pe^{2}/dk_{s}^{2}L^{2}}}
\ = \ \sqrt{\frac{d}{d-1}} \times \Pe^{0}.
\end{equation}

\section{High-$\Pe$ behavior of the mixing efficiency bounds}
\label{sec:SARSAS}

Here we examine the high P\'eclet number behavior of the upper bounds on the
multi-scale mixing efficiencies for statistically homogeneous and isotropic
flows.  As a point of reference we recall that if there is an effective eddy
diffusion associated with the flow field, so that the stirring suppresses the
scalar variances in the manner of enhanced molecular diffusion in the form of
an equivalent (eddy) diffusion $\sim U\ell$, then the efficiencies would scale
``classically'', as ${\cal E}_{p} \sim \Pe^{1}$, as $\Pe \rightarrow \infty$.
Classical scaling allows for the precise identification of equivalent
diffusivities $\kappa^{(\mathrm{eq})}_{p} \ldef \kappa {\cal E}_{p} \rdef U
\ell_{p}$ that serve as the definition of associated ``mixing lengths''
$\ell_{p}$ that are independent of the magnitude of $U$ and $\kappa$.  The
upper limits on ${\cal E}_{0}$ in (\ref{eq:varlubound}), ${\cal E}_{1}$ in
(\ref{E0cheap}) and ${\cal E}_{-1}$ (\ref{eq:igvarlubound}), reproduced
immediately below for reference, all depend on the full structure of the
source-sink distribution:
\begin{eqnarray}
{\cal E}_{1} &\leq& \sqrt{\frac{\sum_{\bm{k}} |\hat{s}_{\bm{k}}|^2/k^2}
{\sum_{\bm{k}} |\hat{s}_{\bm{k}}|^2/(k^2 + \Pe^2 / L^{2})}} 
\label{E0cheap2a} \\
{\cal E}_{0} &\le& \sqrt{\frac{\sum_{\bm{k}} |\hat{s}_{\bm{k}}|^2/k^4}
{\sum_{\bm{k}} |\hat{s}_{\bm{k}}|^2/(k^4 +k^2 \Pe^2 / L^{2}d)}}
\label{eq:varlubounda} \\
{\cal E}_{-1} &\leq& \sqrt{\frac{\sum_{\bm{k}} |\hat{s}_{\bm{k}}|^2/k^6}
{\sum_{\bm{k}} |\hat{s}_{\bm{k}}|^2/(k^6 +k^4 \Pe^2 / L^{2}d
+k^2 \Pe^2 / \lambda^{2} L^{2}d)} },
\label{eq:igvarluboundb}
\end{eqnarray}
The high-$\Pe$ behavior, though, can be discerned from just a few features
of the sources and sinks.

\subsection{Square-integrable sources and sinks}

Consider the case where the source-sink distribution function is square integrable,
$s(\bm{x}) \in L^{2}(\mathbb{T}^d)$, so that the Fourier coefficients are square summable:
\begin{equation}
\sum_{\bm{k}} |\hat{s}_{\bm{k}}|^{2} \ < \ \infty.
\end{equation}
Then the $\Pe \rightarrow \infty$ asymptotic behaviors in (\ref{E0cheap2a}), 
(\ref{eq:varlubounda}) and (\ref{eq:igvarluboundb}) are elementary to evaluate:
\begin{eqnarray}
{\cal E}_{1} \ &\lesssim& \ \Pe \
\sqrt{\frac{\sum_{\bm{k}} |\hat{s}_{\bm{k}}|^2/k^2}
{L^{2}\sum_{\bm{k}} |\hat{s}_{\bm{k}}|^2}} \ \rdef \
\Pe \times \frac{\ell_{1}^{(\mathrm{max})}}{L}
\label{HighPeL2p1}
\\
{\cal E}_{0} \ &\lesssim& \ \Pe \
\sqrt{\frac{\sum_{\bm{k}} |\hat{s}_{\bm{k}}|^2/k^4}
{dL^{2}\sum_{\bm{k}} |\hat{s}_{\bm{k}}|^2/k^2}} \ \rdef \
\Pe \times \frac{\ell_{0}^{(\mathrm{max})}}{L}
\label{HighPeL2v0}
\\
{\cal E}_{-1}  \ &\lesssim& \ \Pe \
\sqrt{\frac{\sum_{\bm{k}} |\hat{s}_{\bm{k}}|^2/k^6}
{dL^{2}\sum_{\bm{k}} |\hat{s}_{\bm{k}}|^2/(k^{4}
+k^2 / \lambda^{2})}} \ \rdef \
\Pe \times \frac{\ell_{-1}^{(\mathrm{max})}}{L}.
\label{HighPeL2m1}
\end{eqnarray}
There are three significant features of these scaling bounds worth noting.

The first point is that these upper estimates all scale classically, allowing
us to identify the largest possible values for meaningful mixing lengths that
we have labeled $\ell_{p}^{(\mathrm{max})}$.

The second remarkable fact is that the largest possible mixing lengths
relevant to small and intermediate scale fluctuations do {\it not} depend on
the flow field, but rather only on the source-sink distribution.  Indeed,
$\ell_{1}^{(\mathrm{max})} \le \ell_{0}^{(\mathrm{max})} \le L/2\pi$ are particular length
scales in the source-sink function that have nothing to do with the stirring
field or any length scales in the flow.  This is in direct conflict with the
notion that a mixing length should be a characteristic persistence length or
eddy size in the velocity vector field.  Rather, these mixing efficiencies are
ultimately limited by the structure of the sources and sinks.  When
$\ell_{1}^{(\mathrm{max})}$ and $\ell_{0}^{(\mathrm{max})}$ are ``small'' (i.e., when
$\ell_{1}^{(\mathrm{max})} \le \ell_{0}^{(\mathrm{max})} \ll L$) then the mixing efficiencies are
also ``small'' and {\it not} subject to any further improvement by any sort of
clever stirring designed to enhance the variance reduction at intermediate and
small scales.

The bound on the mixing length $\ell_{-1}$ for large-scale variance reduction,
however, {\it does} depend on the spectrum of length scales in the flow
through (the Taylor microscale) $\lambda$.  It is interesting to note that the
bound on $\ell_{-1}^{(\mathrm{max})}$ is an increasing function of $\lambda^{-1}$,
allowing for the possibility that small-scale stirring could enhance large
scale mixing.  That is, this analysis does not preclude small-scale stirring from
suppressing large-scale fluctuations in ways that it cannot decrease the variance at
intermediate and small scales.

We note as well that the improved upper bound on the small-scale mixing
efficiency ${\cal E}_{1}$ in (\ref{IB}) generally produces an even {\it
smaller} estimate for $\ell_{1}^{(\mathrm{max})}$ that depends on the spectrum, i.e.,
the magnitude and distribution, of the length scales in the flow.  In cases
where the shortest length scale (call it $\ell_{s}$) in the source is much
longer than the longest length scale (call it $\ell_{u}$) in the stirring
field, a thoughtful examination of (\ref{IB}) suggests that $\ell_{1}^{(\mathrm{max})}
\sim \ell_{u} \ll \ell_{s}$.

The third point worth noting is that the upper estimates for all three of the
${\cal E}_{p}$ scale the same, $\sim \Pe^{1}$ as $\Pe \rightarrow \infty$.  So
far all the examples we have considered share this property, but in the next
subsection we will see that this is {\it not} a general feature of the
multi-scale mixing efficiency bounds.

\subsection{Measure-valued source-sink distributions}

If $s(\bm{x}) \notin L^{2}(\mathbb{T}^d)$ then the sum in the denominator of
(\ref{HighPeL2p1}) diverges and the ratio defining the prefactor
($\ell_{1}^{(\mathrm{max})}/L$) to the $\Pe^{1}$ scaling could vanish.  This
would violate the lower bound ${\cal E}_{1} \ge 1$, so the high-$\Pe$
asymptotic analysis of the ratios of sums must be revisited.  This issue
arises for measure-valued sources and sinks, i.e., when the distribution
involves singular objects like $\delta$-functions.  
Then ``anomalous'' sub-classical P\'eclet number scalings for some
efficiencies are inevitable.

Consider the extreme cases where the Fourier coefficients of $s(\bm{x})$ obey
\begin{equation}
|\hat{s}(\bm{k})| = {\cal O}(1) \quad \text{as} \quad 
|\bm{k}| \rightarrow \infty.
\end{equation}
In this case, in spatial dimensions $d=2$ and $3$, sums in both
(\ref{HighPeL2p1}) and (\ref{HighPeL2v0}) diverge so the high-$\Pe$ behavior
of the bounds on ${\cal E}_{1}$ and ${\cal E}_{0}$ must be re-evaluted
directly from (\ref{E0cheap2a}) and (\ref{eq:varlubounda}).  The classical
high-$\Pe$ scaling (\ref{HighPeL2m1}) for the bound on ${\cal E}_{-1}$ from
(\ref{eq:igvarluboundb}) remains the same in $d=2$ and $3$.

To discern the high P\'eclet number behavior of the expressions in
(\ref{E0cheap2a}) and (\ref{eq:varlubounda}) when $|\hat{s}(\bm{k})| \sim C =
{\cal O}(1)$ as $|\bm{k}| \rightarrow \infty$, we analyze integral
approximations to the sums.  This is justified because the major contribution
to these diverging sums comes from the high-$k$ end where the discreteness
of the wavenumbers is negligible.  For example, the denominator for the bound
on ${\cal E}_{0}$ in (\ref{eq:varlubounda}) is
\begin{equation}
\sum_{\bm{k}} \frac{|\hat{s}_{\bm{k}}|^2}
{k^4 + k^{2}\Pe^2 / L^{2}d}
\ \approx \ \left( \frac{L}{2\pi} \right)^{d} \ S_{d}
\int_{\fracs{2\pi}{L}}^{\infty} \frac{C \ k^{d-3} dk}
{k^{2}+\Pe^2 / L^{2}d}
\end{equation}
where $S_{2} = 2\pi$ and $S_{3}=4\pi$.

In $d=2$ this is
\begin{equation}
\sum_{\bm{k}} \frac{|\hat{s}_{\bm{k}}|^2}
{k^4 + k^{2}\Pe^2 / L^{2}d}
\ \approx \ \frac{L^{2}}{2\pi} \ C \ \frac{L^{2}}{\Pe^{2}}
\ \log \left[1+\frac{\Pe^{2}}{8\pi^{2}} \right],
\end{equation}
so the upper bound on ${\cal E}_{0}$ in $d=2$ is
\begin{equation}
{\cal E}_{0} \ \lesssim \
\frac{\Pe}
{\sqrt{8\pi^{2}\log \left[1+\frac{\Pe^{2}}{8\pi^{2}} \right]}}
\ \sim \ \frac{\Pe}{(\log{\Pe})^{1/2}}
\quad \text{as} \ \Pe \rightarrow \infty.
\end{equation}
This upper bound exhibits a logarithmic correction to classical scaling.  The
significance of this is perhaps best appreciated as the {\it absence} of any
residual variance suppression in the limit of vanishing molecular diffusivity.
That is, the largest possible effective diffusivity defined by the bulk
variance suppression $\kappa_{\mathrm{eff}} \le \kappa {\cal E}_{0} \sim {\cal
O}(|\log{\kappa}|^{-1/2}) \rightarrow 0$ as $\kappa \rightarrow 0$.  It is
worthwhile stressing that this result does not depend on---and cannot be
circumvented by manipulating---any further details of the statistically
homogeneous and isotropic stirring field.

The deviation from classical scaling is more dramatic in $d=3$ where
\begin{equation}
\sum_{\bm{k}} \frac{|\hat{s}_{\bm{k}}|^2}
{k^4 + k^{2}\Pe^2 / L^{2}d}
\ \approx \ \left( \frac{L}{2\pi} \right)^{3} \ 4 \pi C \ \frac{\sqrt{3}L}{\Pe}
\ \left[ \frac{\pi}{2} - \arctan \left(\frac{2\pi\sqrt{3}}{\Pe} \right) \right].
\end{equation}
Then the upper bound on ${\cal E}_{0}$ is
\begin{equation}
{\cal E}_{0} \ \lesssim \  
\sqrt{
\frac{\Pe}{
2 \pi \sqrt{3} \ \left[ \frac{\pi}{2} 
- \arctan \left(\frac{2\pi\sqrt{3}}{\Pe} \right)\right]
}
} 
\ \sim \ \Pe^{1/2}
\quad \text{as} \ \Pe \rightarrow \infty.
\end{equation}
This upper bound exhibits strictly
sub-classical scaling at high P\'eclet numbers.

Both the numerator and the denominator diverge in the bound in
(\ref{E0cheap2a}) for ${\cal E}_{1}$ for such measure-valued source-sink
distributions, so those bounds must be evaluated as the limit of ratios for a
sequence of appropriately mollified sources and sinks.  This is
straightforward if we simply truncate the Fourier series for $s(\bm{x})$ at
small length scale $\ell_{s} \ll L$ and study the ideal case where
$|\hat{s}(\bm{k})| = C$ for $2\pi/L \le |\bm{k}| < 2\pi/\ell_{s}$ and
$|\hat{s}(\bm{k})| = 0$ for $|\bm{k}| > 2\pi/\ell_{s}$, and then take the
limit $\ell_{s}/L \rightarrow 0$.

Again we analyze integral approximations to the sums;
the numerator in (\ref{E0cheap2a}) is
\begin{equation}
\sum_{\bm{k}} \frac{|\hat{s}_{\bm{k}}|^2}{k^2}
\ \approx \ \left( \frac{L}{2\pi} \right)^{d} \ S_{d}
\int_{\fracs{2\pi}{L}}^{\fracs{2\pi}{\ell_{s}}} \frac{C \ k^{d-1} dk}
{k^{2}}
\end{equation}
and the denominator is
\begin{equation}
\sum_{\bm{k}} \frac{|\hat{s}_{\bm{k}}|^2}{k^2 + \Pe^2 / L^{2}}
\ \approx \ \left( \frac{L}{2\pi} \right)^{d} \ S_{d}
\int_{\fracs{2\pi}{L}}^{\fracs{2\pi}{\ell_{s}}} \frac{C \ k^{d-1} dk}
{k^{2}+\Pe^2 / L^{2}}.
\end{equation}
In $d=2$ spatial dimensions this implies the bound
\begin{equation}
{\cal E}_{1} \ \lesssim \  
\sqrt{
\log \left[ \frac{L^{2}}{\ell_{s}^{2}}\right]
} \bigg/
\sqrt{
\log \left[ \frac{\Pe^{2}+4 \pi^{2} \frac{L^{2}}{\ell_{s}^{2}}}
{\Pe^{2}+4 \pi^{2}}\right]
} 
\ \ \rightarrow \ \ 1 
\quad \text{as} \ \ \frac{\ell_{s}}{L} \rightarrow 0.
\end{equation}
Hence there can be {\it no} reduction of the small-scale variance beyond that
due to molecular diffusion for such singular source-sink distributions, no
matter what flow strategy is adopted or how energetic the stirring is.

It is interesting to note as well that if $\ell_{s}$ is small but finite and
$\Pe \gg \frac{L}{\ell_{s}} \gg 1$, then the 
bound scales classically again.
That is, as $\Pe \rightarrow \infty$,
\begin{equation}
{\cal E}_{1} \ \lesssim \  
\sqrt{
\log \left[ \frac{L^{2}}{\ell_{s}^{2}}\right]
}  \bigg/
\sqrt{
\log \left[ \frac{\Pe^{2}+4 \pi^{2} \frac{L^{2}}{\ell_{s}^{2}}}
{\Pe^{2}+4 \pi^{2}}\right]
} 
\ \rightarrow \ \frac{\ell_{s}}{2 \pi \sqrt{L^{2}-\ell_{s}^{2}}} \, \Pe 
\ \approx \ \frac{\ell_{s}}{2 \pi L} \, \Pe.
\end{equation}
This can be interpreted as the statement
that the largest possible value for the mixing length $\ell_{1}^{(\mathrm{max})} =
{\cal O}(\ell_{s})$ in such cases.

For such distributions in $d=3$ spatial dimensions,
\begin{equation}
{\cal E}_{1} \ \lesssim \  
\l({
1 - \frac{\Pe}{\frac{L}{\ell_{s}}-1} \, \left[ 
\arctan\l(\frac{2 \pi L}{\Pe\, \ell_{s}}\r)
- \arctan\l(\frac{2 \pi}{\Pe}\r) \right]
}\r)^{-1/2}
\ \ \rightarrow \ \ 1 
\quad \text{as} \ \ \frac{\ell_{s}}{L} \rightarrow 0.
\end{equation}
Again, there can be no reduction of the small-scale variance beyond that due to
molecular diffusion for this kind of singular source-sink distributions, 
no matter what stirring strategy is adopted.
Furthermore, when $\ell_{s}$ is finite and $\Pe \gg \frac{L}{\ell_{s}} \gg 1$, 
the bound scales classically again: as $\Pe \rightarrow \infty$,
\begin{equation}
{\cal E}_{1} \ \lesssim \  
\l(
{
1 - \frac{\Pe}{\frac{L}{\ell_{s}}-1} \, \left[ 
\arctan\l(\frac{2 \pi L}{\Pe \ell_{s}}\r)
- \arctan\l(\frac{2 \pi}{\Pe}\r) \right]
}\r)^{-1/2}
\ \rightarrow \ 
\sqrt{
\frac{3}{4\pi^{2}} \, 
\frac{\frac{L}{\ell_{s}}-1}{\frac{L^{3}}{\ell_{s}^{3}}-1}
}
 \, \Pe 
\ \approx \ \frac{\sqrt{3} \, \ell_{s}}{2 \pi L} \, \Pe.
\end{equation}
Not unexpectedly, the largest possible
value for a mixing length for this kind of mollified distribution is
$\ell_{1}^{(\mathrm{max})} = {\cal O}(\ell_{s})$.

Classical scaling for ${\cal E}_{0}$ is also eventually recovered for these
regularlized sources and sinks when $\Pe \gg L/\ell_{s}$, albeit with very
different relationships between the smallest source-sink length scale
$\ell_{s}$ and the maximal mixing length $\ell_{0}^{(\mathrm{max})}$.  When
$|\hat{s}(\bm{k})| = C > 0$ for $2\pi/L \le |\bm{k}| < 2\pi/\ell_{s}$ and
$|\hat{s}(\bm{k})| = 0$ for $|\bm{k}| > 2\pi/\ell_{s}$ the numerator of the
bound in (\ref{eq:varlubounda}) is
\begin{equation}
\sum_{\bm{k}} \frac{|\hat{s}_{\bm{k}}|^2}{k^4}
\ \approx \ \left( \frac{L}{2\pi} \right)^{d} \ S_{d}
\int_{\fracs{2\pi}{L}}^{\fracs{2\pi}{\ell_{s}}} \frac{C \ k^{d-1} dk}
{k^{4}}
\end{equation}
and the denominator is
\begin{equation}
\sum_{\bm{k}}\frac {|\hat{s}_{\bm{k}}|^2}{k^4 +k^2 \Pe^2 / L^{2}d}
\ \approx \ \left( \frac{L}{2\pi} \right)^{d} \ S_{d}
\int_{\fracs{2\pi}{L}}^{\fracs{2\pi}{\ell_{s}}} \frac{C \ k^{d-1} dk}
{k^{4}+k^{2}\Pe^2 / L^{2}d}.
\end{equation}
For $d=2$, as $\Pe \rightarrow \infty$,
\begin{equation}
{\cal E}_{0} \ \lesssim \
\sqrt{
\frac{\Pe^{2}}{8 \pi^{2}}
\left( 1 - \frac{\ell_{s}^{2}}{L^{2}} \right)
\frac{1}{\log \left[
\frac{8\pi^{2}+\Pe^{2}}{8\pi^{2}+\ell_{s}^{2}\Pe^{2}/L^{2}}
\right]}
}
\ \ \sim \ \ 
\frac{\Pe}{2 \pi \sqrt{2}} \sqrt{
\frac{\left( 1 - \frac{\ell_{s}^{2}}{L^{2}} \right)}
{\log \frac{L^{2}}{\ell_{s}^{2}}}
}
\ \ \approx \ \  
\frac{\Pe}{4 \pi}
\sqrt{\frac{1}{\log \frac{L}{\ell_{s}}}} \ .
\end{equation}
So in spatial dimension $d=2$ the largest possible mixing length
$\ell_{0}^{(\mathrm{max})} \sim L/\sqrt{\log[L/\ell_{s}]}$.  This length scale
vanishes as $\ell_{s} \rightarrow 0$, but much slower than the actual smallest
scale ${\cal O}(\ell_{s})$.

In $d=3$,
\begin{equation}
{\cal E}_{0} \ \lesssim \
\frac{\Pe}{2 \pi \sqrt{2}}\sqrt{
\frac{\ell_{s}}{L}
}
\quad \text{as} \ \Pe \rightarrow \infty
\end{equation}
so $\ell_{0}^{(\mathrm{max})} = {\cal O}(\ell_{s}^{1/2})$.
This vanishes faster than in $2$-$d$ as $\ell_{s} \rightarrow 0$, but
still much slower than $\ell_{s}$.
 
\begin{figure}
\centerline{\includegraphics[width=10.0cm]{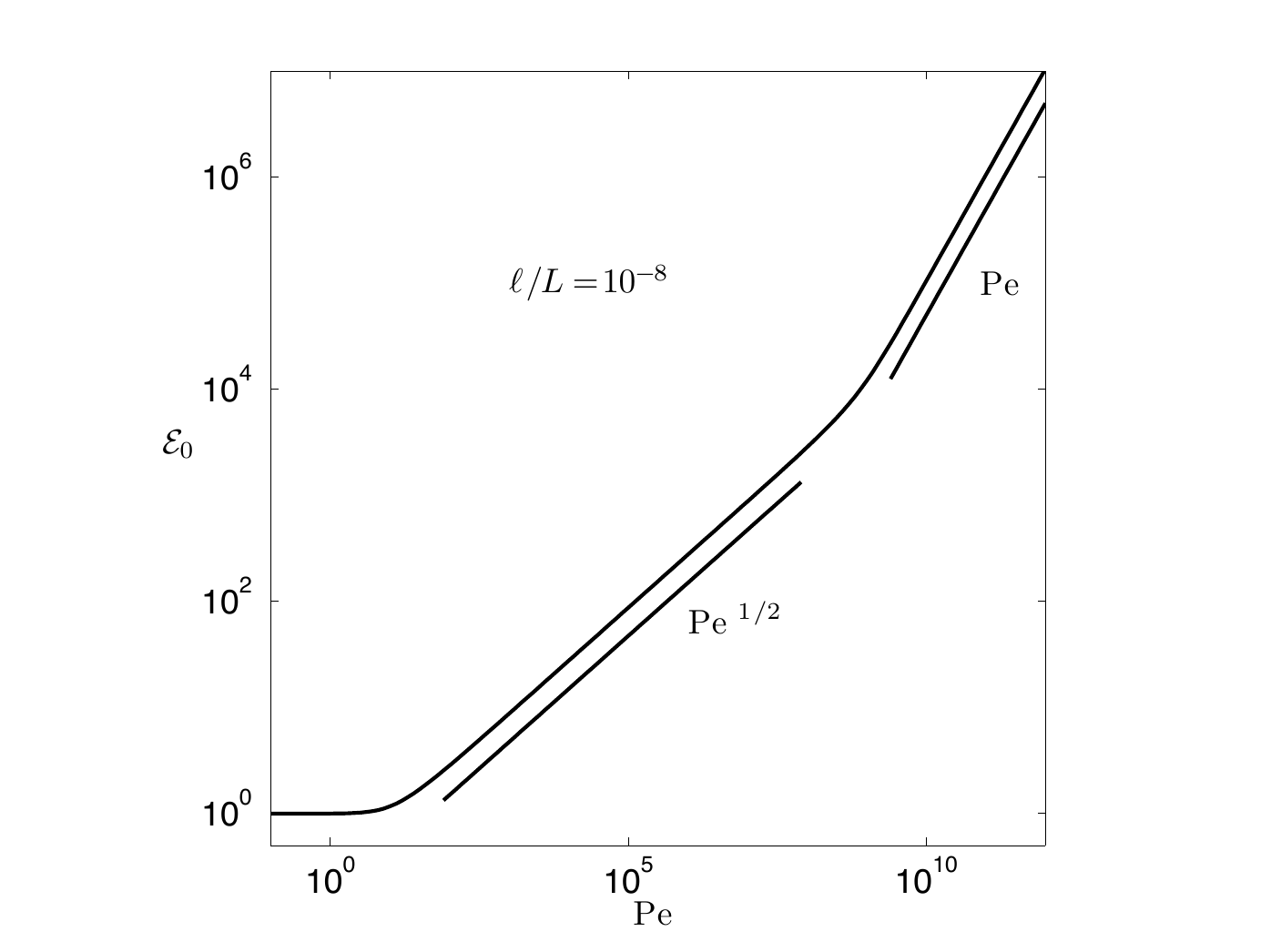}}
\setlength{\abovecaptionskip}{-5pt}
\caption{${\cal E}_0$ as a function of $\Pe$ for cutoff source-sink
distributions, after~\cite{DoeringThiffeault2006}.}
\label{dumfig1}
\end{figure}

A log-log plot of the bound on ${\cal E}_{0}$ vs $\Pe$ for these cutoff
source-sink distributions is shown in Figure~\ref{dumfig1} for the
$3$-$d$ case with $\ell_{s}/L = 10^{-8}$.  The anomalous $\Pe^{1/2}$ scaling
persists over the range $1 \lesssim \Pe \lesssim L/\ell_{s}$ and classical
scaling $\sim \Pe^{1}$ takes over at higher P\'eclet numbers with a ``small''
prefactor $\sim \sqrt{\ell_{s}/L}$.  This small prefactor is a quantitative
indication of the difficulty of any statistically homogeneous and isotropic
flow field to efficiently suppress the scalar variance---beyond the suppression
achieved by molecular diffusion alone---in the presence of small-scale scalar
sources and sinks.

The scalings for the mixing efficiencies with a $\delta$-function source are 
apparently not all realized by
a simple uniform wind as was the case for smooth monochromatic sources.
We have performed the adiabatic approximation for the bulk variance 
in the case of a uniform wind blowing past a $\delta$-function source
to estimate the mixing efficiencies (see Appendix B).
While ${\cal E}_1$ is necessarily equal to 1 in both $2$-$d$ and $3$-$d$,
we find that the uniform wind gives ${\cal E}_0 \sim \sqrt{\Pe}$ (rather than
the bound $\Pe/\sqrt{\log{\Pe}}$) in $2$-$d$, and 
${\cal E}_0 \sim \sqrt{\Pe}/\sqrt{\log{\Pe}}$ 
(significantly closer to the bound $\sqrt{\Pe}$) in $3$-$d$.
For the large-scale efficiency, however, the uniform wind produces 
${\cal E}_{-1} \sim \sqrt{\Pe}$ in both $2$-$d$ and $3$-$d$,
far below its classically scaling upper bound.

A similar integral analysis can be carried out for ``fractal'' source-sink
distributions where $|\hat{s}_{\bm{k}}| \sim |\bm{k}|^{-\gamma}$ with $\gamma
\ge 0$.  The scalings of the mixing efficiency bounds are summarized in
Table~\ref{tab:scalingsall}.  For $\gamma > d/2$ all the bounds scale
classically since then $s(\bm{x}) \in L^{2}(\mathbb{T}^d)$.  The
case~$\gamma=0$ is the $\delta$-function studied above.  We observe that the
bounds for ${\cal E}_{1}$ and ${\cal E}_{0}$ can exhibit anomalous scaling
with exponents depending on the fractal nature of the source-sink distribution
as characterized by $\gamma$~\cite{DoeringThiffeault2006}.  Of course if the
fractal scaling of the source-sink distribution persists over a broad but
finite range of wavenumbers, say for $2\pi/L < k < 2\pi/\ell_{s}$, then
classical scaling for the efficiency estimates will again appear for $\Pe \gg
L/\ell_{s}$.

\begin{table}
\vspace{.5em}
\caption{Scalings of the bound on the mixing efficiency~${\cal E}_{p}$ as
functions of the source roughness exponent~$\gamma$ of the source in two and
three dimensions.}
\begin{flushleft}
\begin{ruledtabular}
\begin{tabular}{lccc}
$d = 2$ & ${p}=1$ & ${p}=0$ & ${p}=-1$ \\
\hline
$\gamma = 0$       & 1 & $\Perm/(\log\Perm)^{1/2}$ & \Perm \\
$0 < \gamma < 1$ & $\Perm^\gamma$ & \Perm & \Perm \\
%$\gamma = 1/2$     & $\Pe^{1/2}$ & \Pe  & \Pe \\
%$1/2 < \gamma < 1$ & $\Pe^\gamma$ & \Pe & \Pe \\
$\gamma = 1$       & \ \ \ $\Perm/(\log\Perm)^{1/2}$ & \Perm & \Perm \\
$\gamma > 1$       & \Perm & \Perm & \Perm \\[2pt]
%\end{tabular}
%\end{ruledtabular}
%\vspace{1.0em}
%\begin{ruledtabular}
%\begin{tabular}{lccc}
\hline
$d = 3$\\ %& ${p}=1$ & ${p}=0$ & ${p}=-1$ \\
\hline
$\gamma = 0$         & 1 & $\Perm^{1/2}$ & \Perm \\
$0 < \gamma < 1/2$   & 1 & $\Perm^{\gamma+1/2}$ & \Perm \\
$\gamma = 1/2$       & 1 & \quad $\Perm/(\log{\Perm})^{1/2}$ & \Perm \\
$1/2 < \gamma < 3/2$ & $\Perm^{\gamma - 1/2}$ & \Perm & \Perm \\
$\gamma = 3/2$       & \ $\Perm/(\log{\Perm})^{1/2}$ & \Perm & \Perm \\
$\gamma > 3/2$       & \Perm & \Perm & \Perm \\
\end{tabular}
\end{ruledtabular}
\end{flushleft}
\label{tab:scalingsall}
\end{table}

%===================================================

\section{A Single-scale Source Stirred by a Single-scale Flow}
\label{sec:SSSSSF}

In the previous sections we have seen that the classical high-P\'eclet number
scalings for the multi-scale mixing efficiencies, ${\cal E}_{p} \sim \Pe$, are
generally upper bounds that may in fact be saturated for particular
source-sink distributions stirred by certain statistically homogeneous and
isotropic flows.  It was shown in Section~\ref{sec:SARSAS} that for some
distribution-valued sources and sinks, some of the mixing efficiencies
necessarily scale anomalously, i.e., ${\cal E}_{p} \lesssim \Pe^{\alpha_{p}}$
with some $\alpha_{p} < 1$.  For those examples the bounds on the mixing
efficiencies at the different scales scale differently: $0 \le \alpha_{1} <
\alpha_{0} < \alpha_{-1} = 1$.  It is not presently known if the anomalously
scaling bounds for distribution-valued sources and sinks are sharp or, if they
are, what kinds of statistically homogeneous and isotropic flows might be
required to realize them.  Those questions remain open, but in this section we
settle the issue of the possibility of realizing distinct scaling exponents
for the mixing efficiencies on different length scales for a smooth
source-sink distribution.

The random sine flow, a.k.a.\ the renewing wave flow characterized by a single
length scale, is a popular and convenient test bed for studying---both
analytically and via simulation---a wide variety of stirring and mixing
phenomena~\cite{Pierrehumbert1994,Antonsen1996,Thiffeault2004,Tsang2005}.
Here we consider the simplest single-scale source
\begin{equation}
s(\bm{x}) = \sqrt{2} \, S \sin[2 \pi (x+y)/L]
\end{equation}
stirred by a random sine flow that switches between
\begin{equation}
\bm{u}(\bm{x}) = \hat{\bm{i}} \sqrt{2} \, U \sin[2 \pi y/L + \phi]
\label{Z1}
\end{equation}
and
\begin{equation}
\bm{u}(\bm{x}) = \hat{\bm{j}} \sqrt{2} \, U \sin[2 \pi x/L + \phi]
\label{Z2}
\end{equation}
at time intervals of length $\tau/2$, where the phase $\phi$ is chosen
randomly and uniformly from $[0, 2\pi)$ at each switch.  This is a
statistically homogeneous and isotropic flow, but it is not the maximally
efficient flow for this source-sink distribution; the optimal flow is the
spatially uniform flow constructed in Section~\ref{sec:saturation}.  We have
performed high resolution direct numerical simulations of this system in $d=2$
spatial dimensions that reveal that the multi-scale mixing efficiencies may
have distinct scaling exponents.  Additionally, for this example we are able
to compute the exponents theoretically in the context of the adiabatic
averaging approximation utilized in Section~\ref{sec:saturation}.

\begin{figure}
\centerline{\includegraphics[width=19cm]{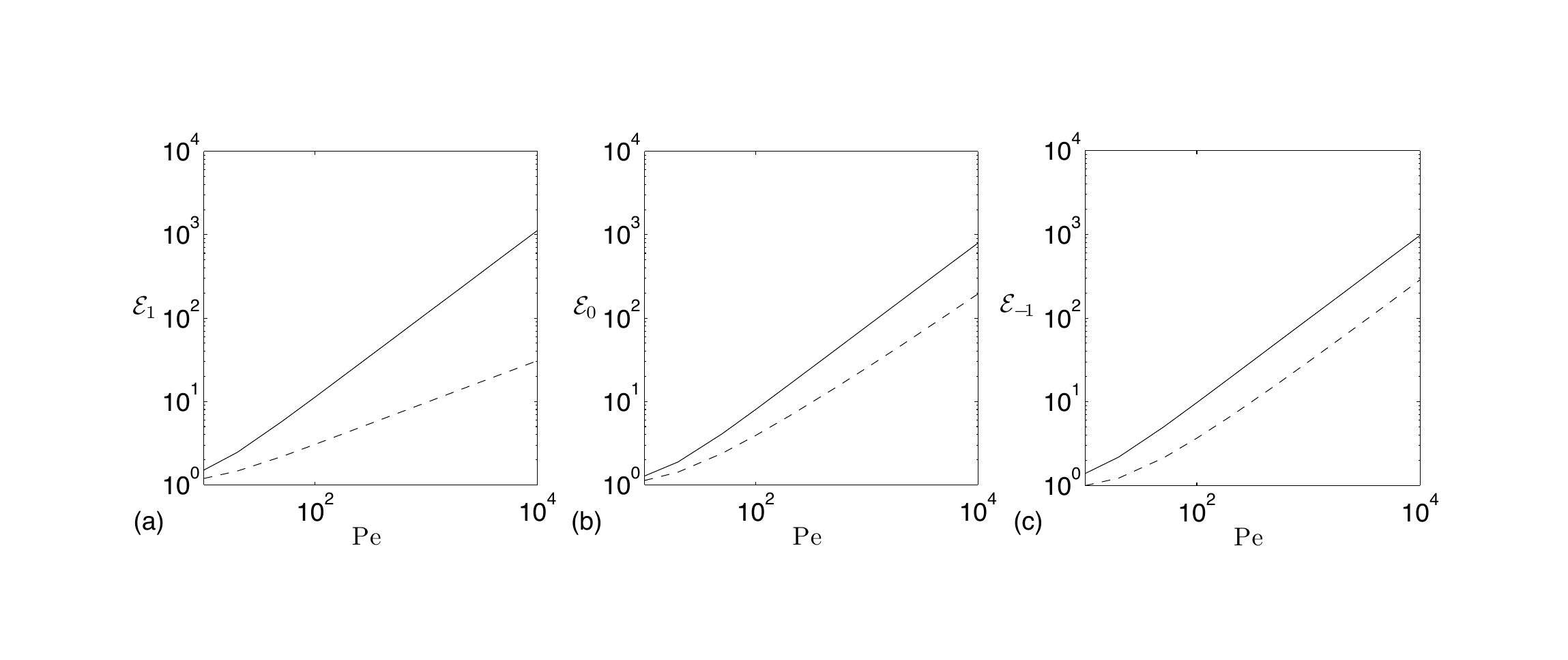}}
\setlength{\abovecaptionskip}{-30pt}
\caption{ Mixing efficiencies ${\cal E}_p$ as a function of $\Pe$ for (a)
$p=1$, (b) $p=0$, and (c) $p=-1$ for the random sine flow with source $\sim
\sin k_s(x+y)$.  The solid lines are the upper bounds for this source from
Section~\ref{sec:BOUNDS} and the dashed lines are the data from the direct
numerical simulations with $U$, $L$ and $\tau$ fixed.}
\label{dumfig2}
\end{figure}

Figure~\ref{dumfig2} shows the results of the direct numerical simulation for
$p=1,0,-1$, each plotted along with the classically-scaling upper bounds from
Section~\ref{sec:BOUNDS}.  The P\'eclet number in these simulations was varied
by holding $U=1$ fixed in a box of side length $L=1$ 
and changing the molecular diffusivity $\kappa$.  
The computations were carried out with switching time $\tau=1$.  
The details of the numerical simulations can be found
in~\cite{Thiffeault2004}.
Writing ${\cal E}_{p} \sim \Pe^{\alpha_{p}}$ as $\Pe \rightarrow \infty$,
it is clear that the data are 
consistent with $\alpha_{0} < \alpha_{1} < 1$ and $\alpha_{-1} \approx 1$.

These simulations establish two important facts:
(1) that the mixing efficiencies at different length scales generally scale differently at
high P\'eclet numbers, and (2) that anomalous subclassical scaling can easily be 
realized by simple and ``reasonable'' flows.

We can evaluate the observed scaling exponents $\alpha_{p}$ theoretically for
this particular system by appealing to a quasi-static adiabatic approximation
introduced in Section~\ref{sec:saturation} for the optimally mixing uniform
flow.  That is, we solve the time-independent problem for the scalar
distribution under the influence of steady flows of the form (\ref{Z1}) and
(\ref{Z2}), compute the variances, and average over the flow configurations.
The physical justification for this approximation comes from examining
the results of the direct numerical simulations: it is observed that the
relevant steady-state configuration is quickly approached in the time
intervals between the switches of the stirring field.  If we make the
switches at longer and longer intervals, i.e., if we increase $\tau$,
the scalar field is distributed (nearly) according to the static
configurations for most of the time, and the steady-state variances
dominates the time averages.\footnote{This approach is similar in spirit
to ``rapid distortion theory'' in turbulence~\cite{Savill1987,Hunt1990}.}

The simplest problem in this category is when the velocity field is oriented
parallel to the gradient of the source, so we consider the steady
advection--diffusion equation
\begin{equation}
  \sqrt{2} U \sin \ku y \  \partial_{x} \theta =
  \kappa (\partial_{x}^{2}+\partial_{y}^{2}) \theta + 
  \sqrt{2} S \sin \ks x
  \label{eq:ADsteady}
\end{equation}
where we allow for different length scales in the stirring and the source with
the non-dimensional number $r=\ku/\ks$ gauging the relative amount of shear in
the flow.  This is not exactly the static problem corresponding to the
dynamic simulations where the flow is always at a $45\degree$ angle from the
source-sink alignment, but it turns out that the scaling exponents
$\alpha_{p}$ are the same for that case and for the $90\degree$ alignment in
(\ref{eq:ADsteady}).  For illustrative purposes it is convenient to explain
the most elementary case (\ref{eq:ADsteady}) in detail here
and relegate details of the $45\degree$ problem to Appendix C.

The solution to~\eqref{eq:ADsteady} takes the form
\begin{equation}
  \theta(\bm{x}) = f(y)\sin(\ks x) + g(y) \cos(\ks x)
\end{equation}
where the functions $f$ and $g$ are periodic on $y\in[-\pi/\ku,\pi/\ku]$ 
and satisfy the system of ODEs
\begin{subequations}
\begin{align}
  -\sqrt{2}U\ks \sin(\ku y)\, g(y)
  &= \kappa \left [ -\ks^2 + \frac{d^2}{dy^2}
    \right] f(y) + \sqrt{2} S\,,\label{54a}\\
  \sqrt{2}U\ks \sin(\ku y) f(y) &= \kappa \left [ -\ks^2 + \frac{d^2}{dy^2}
    \right] g(y)\,.\label{54b}
\end{align}
\label{eq:fgeqns}%
\end{subequations}%
From~\eqref{54a} we deduce that $g(y)$ is an odd function of~$y$ and~$f(y)$ is
an even function of~$y$, compatible with~\eqref{54b} which says
that~$f$ and~$g$ have opposite parity.
We can thus infer boundary conditions on the reduced domain $[0,\pi/\ku]$:
\begin{equation}
  g(0)=0=g(\pi/\ku),\quad f'(0)=0=f'(\pi/\ku).
\end{equation}

\begin{figure}
\centerline{\includegraphics[width=20.0cm]{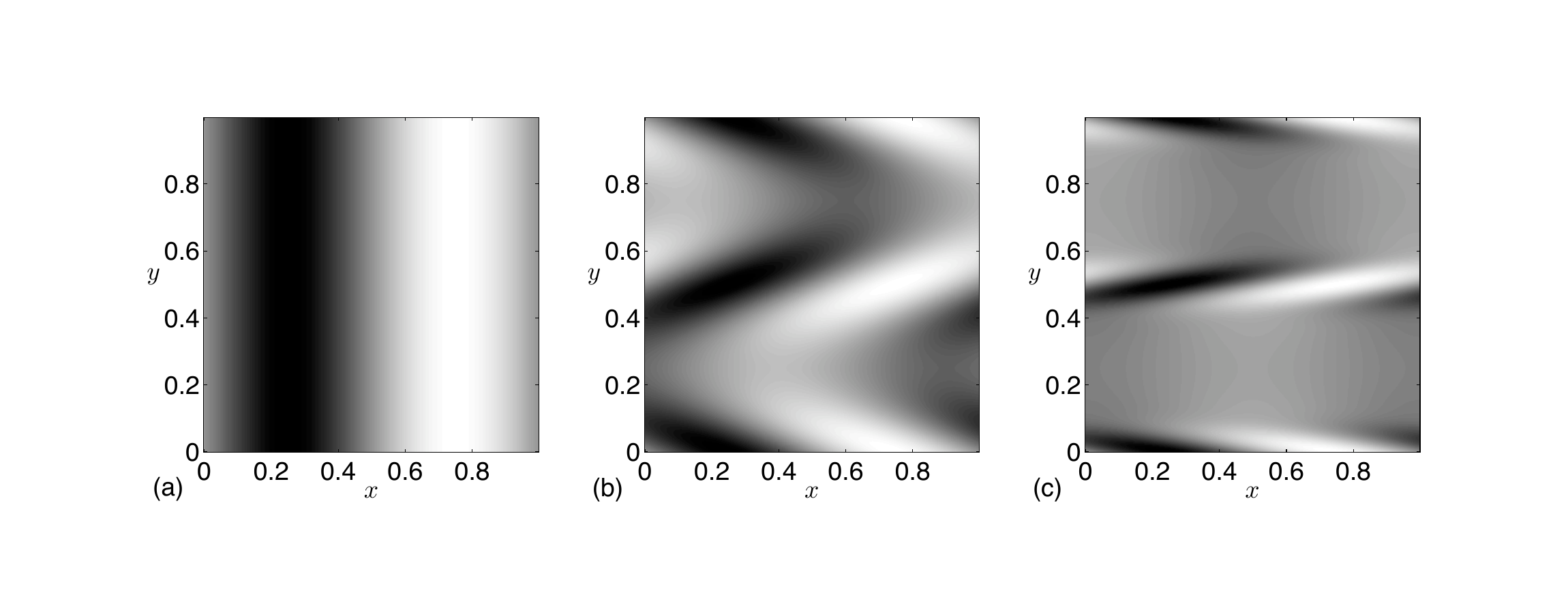}}
\setlength{\abovecaptionskip}{-25pt}
\caption{The scalar field stirred by a steady sine flow for (a) $\Pe$ = 0, (b) $\Pe$ = 100,
 (c) $\Pe$ = 1000.}
\label{dumfig3}
\end{figure}

We are interested in the high-$\Pe$ behavior of the solution
to~\eqref{eq:ADsteady}.  In Figure~\ref{dumfig3} we show a grayscale plot of
the scalar field at several values of the P\'eclet number.  As is evident,
internal layers develop along the lines of maximum shear around $y = n
\pi/\ku$ for $n=0, \pm1, \pm2, \dots$.  Away from these lines the flow
directly blows source regions onto sink regions and vice versa which, as we
have seen in Section~\ref{sec:saturation}, is the most efficient stirring
to reduce variance at all scales.  Even though the regions of high shear
effectively stretch material lines, the mixing process is apparently
frustrated by the constant replishment of the scalar variation by the steady
sources and sinks and the variance is dominated by fluctuations concentrated
in shear layers.

We develop an asymptotic singular perturbation internal layer approach for the
limit $\Pe \gg 1$ with~$r = {\cal O}(1)$ fixed.  Upon rescaling $\tilde{y}=\ks
y$, $\hat{f}=fU\ks/S$, $\hat{g}=gU\ks/S$, $r=\ku/\ks$, and introducing the
slightly re-scaled P\'eclet number $\Pec=\sqrt{2}U/\kappa \ks = \sqrt{2}
\Pe/\ks L$ in Eqs.~\eqref{eq:fgeqns}, we obtain the non-dimensional ODEs
\begin{subequations}
\begin{align}
  \frac{1}{\Pec}  \left[ -1 + \frac{d^2}{d \tilde{y}^2} \right]
  \hat{f}(\tilde{y})+1 &= -\sin(r\tilde y) \hat{g}(\tilde{y}),\label{55a}\\
  \frac{1}{\Pec}  \left[ -1 + \frac{d^2}{d \tilde{y}^2} \right]
  \hat{g}(\tilde{y}) &= \sin(r\tilde y) \hat{f}(\tilde{y}).\label{55b}
\end{align}
\label{eq:fgeqnsscaled}%
\end{subequations}%
Proceeding as usual we construct inner and outer solutions.  The outer
solution, valid away from the internal layers, is obtained by expanding in
powers of $\Pec^{-1}$:
\begin{equation}
  \hat{f}_{\mathrm{outer}}=\sum_{n=0}^{\infty}  \Pec^{-n}
  \hat{f}_{n},\qquad\hat{g}_{\mathrm{outer}}=\sum_{n=0}^{\infty} \Pec^{-n}
  \hat{g}_{n}.
\end{equation}
To leading order the solution to~\eqref{eq:fgeqnsscaled} in the outer region is
\begin{equation}
  \hat{f}_{\mathrm{outer}} \sim 0,\qquad 
  \hat{g}_{\mathrm{outer}} \sim -\frac{1}{ \sin(r \tilde{y})}\,.
  \label{eq:outersol}
\end{equation}

For the internal layer we expand in a small parameter~$\epsilon$ as
\begin{equation}
  \hat{f}_{\mathrm{inner}}= \sum_{n=-1}^{\infty}  \epsilon^n
  \hat{f}_{n},\qquad \hat{g}_{\mathrm{inner}}=
  \sum_{n=-1}^{\infty}  \epsilon^n \hat{g}_{n}\,,
  \label{eq:innersol}
\end{equation}
so that both~$\hat{f}_{\mathrm{inner}}$ and~$\hat{g}_{\mathrm{inner}}$ are
${\cal O}(\epsilon^{-1})$ as~\hbox{$\epsilon\rightarrow 0$}.  The internal
layer scaling is determined by a dominant balance argument: we choose
$\epsilon= \Pec^{-1/3}$ and rescale~$\tilde{y} = \epsilon\, \eta$ to achieve a
self-consistent scaling of the leading order terms.  When these scalings
and~\eqref{eq:innersol} are inserted into~\eqref{eq:fgeqnsscaled}, the problem
to solve at leading order ${\cal{O}}(1/\epsilon)$ is
\begin{equation}
  \frac{d^2 \hat{f}_{-1}}{d \eta^2} + r \eta \hat{g}_{-1} +1=0,
  \qquad\frac{d^2 \hat{g}_{-1}}{d \eta^2} - r \eta \hat{f}_{-1} =0.
  \label{eq:fginner}
\end{equation}
Then letting $\xi=r^{1/3} \eta$, $F=r^{2/3} \hat{f}_{-1}$, and
$G=r^{2/3}\hat{g}_{-1}$, \eqref{eq:fginner} simplifies to
the inner layer equations
\begin{equation}
  F^{''} + \xi G +1 =0,\qquad G^{''} - \xi F = 0,
  \label{510}
\end{equation}
with boundary conditions
%\begin{equation}
  $F'(0)=0$ and $G(0)=0$.
%\end{equation}
The other boundary conditions come from the requirement of matching to the
outer solution~\eqref{eq:outersol}: $F(\xi) \rightarrow 0$ and $ G(\xi)
\rightarrow -1/\xi$ as $\xi \rightarrow \infty$.
The system (\ref{510}) can be cast as a complex Airy equation for $F+iG$, 
but we resort instead to a numerical shooting method to obtain the solution.  

\begin{figure}
\centerline{\includegraphics[width=15cm]{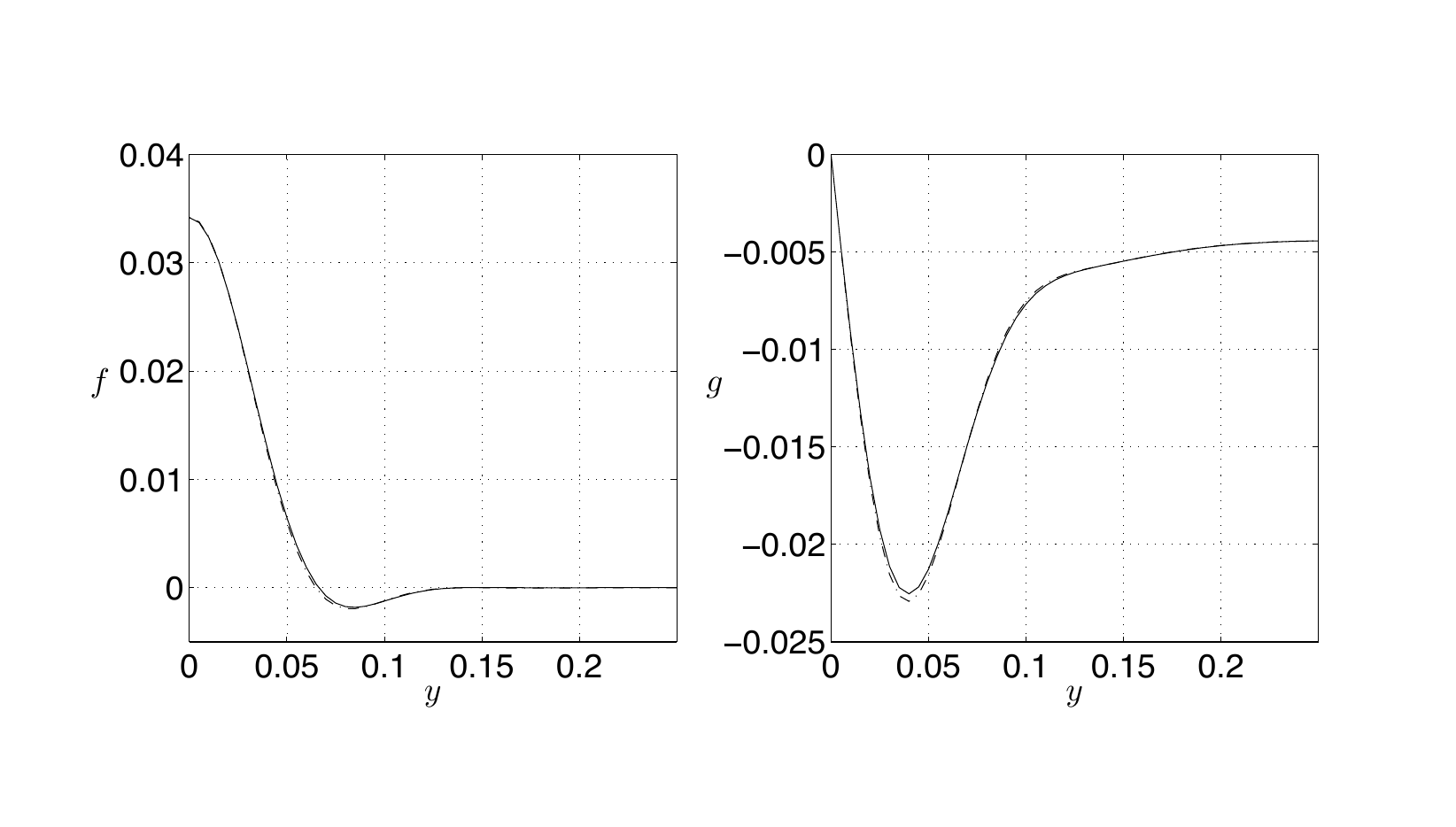}}
\setlength{\abovecaptionskip}{-35pt}
\caption{Comparison of the direct numerical solution (solid) and the internal
layer solution (dashed) for \hbox{$\Pe$ = 1000}.}
\label{dumfig4}
\end{figure}

The solution of the internal layer equations is obtained by shooting backward 
(which is the more stable evolution direction) from $\xi \rightarrow \infty$.
The large-$\xi$ asymptotic behavior deduced from \eqref{510} is
\begin{equation}
  F \approx -\frac{2}{\xi^4} + \frac{a}{\xi^{10}}\,,\qquad
  G \approx -\frac{1}{\xi} + \frac{b}{\xi^7}\,,
\end{equation}
where in practice $a$ and $b$ are tuned numerically to produce a solution that
satisfies the boundary conditions at $\xi=0$.  This is easily accomplished,
and in Figure~\ref{dumfig4} we compare the internal layer solution against the
exact (numerical) solution of the full advection--diffusion equation
(\ref{eq:ADsteady}) at $\Pe=1000$ where the small parameter $\epsilon \approx
.2$.  This agreement confirms that the leading terms
\begin{equation}
  \hat{f}_{\mathrm{inner}} = \frac{r^{-2/3}}{\epsilon}\,F(\xi),
  \qquad
  \hat{g}_{\mathrm{inner}} = \frac{r^{-2/3}}{\epsilon}\,G(\xi).
\end{equation}
do indeed accurately capture the asymptotic behavior.

The uniform asymptotic solution to the coupled ODEs are, to leading order, the
composites of the inner and outer solutions.  Recovering all the scalings and
letting $\delta=\epsilon/r^{1/3} \ks$, we define
\begin{subequations}
\begin{align}
  f_{\mathrm{comp}}(y) \ &= \ \frac{S}{U\ks} \,\hat{f}_{\mathrm{inner}}(y)
  %= \frac{S}{U\ks}
  %\frac{r^{-2/3}}{\epsilon}F \left( \frac{r^{1/3} \ks}{\epsilon} y \right)
  \ = \frac{S}{U\ks}\, \frac{1}{\ku \delta} F \left(\frac{y}{\delta}\right),\\
  g_{\mathrm{comp}}(y) \ &= \ \frac{S}{U\ks} \,
  \ku y \times \hat{g}_{\mathrm{inner}}(y) \times \hat{g}_{\mathrm{outer}}(y)
  %=\frac{S}{U\ks} \frac{r^{-2/3}}{\epsilon}G
  %\left( \frac{r^{1/3} \ks}{\epsilon} y \right) \frac{\ku y}{\sin(\ku y)}
  %\\
  \ = \frac{S}{U\ks}\, \frac{1}{\ku \delta}\,
  \frac{\ku y}{\sin(\ku y)}\,G\left(\frac{y}{\delta}\right).
\end{align}
\label{eq:fginternalsol}
\end{subequations}
Armed with the approximate solutions~\eqref{eq:fginternalsol} we can compute the multi-scale
mixing measures $\langle |\nabla^{p} \theta |^2 \rangle$ for $p=0,1,-1$.

The variance is
\begin{equation}
  \langle \theta^2 \rangle = \frac{1}{2} \left( \langle f^2 \rangle + \langle
  g^2 \rangle \right) =\frac{1}{2} \frac{4\ku}{\pi}
  \left( \int_0^{\pi/4\ku} f(y)^2 dy +
  \int_0^{\pi/4\ku} g(y)^2 dy \right)
\end{equation}
where we use the symmetry of the solution to carry out the integral over only
a quarter-period.  
Letting $\eta=y/\delta$ and replacing $f$ and $g$ by $f_{\mathrm{comp}}$ and $g_{\mathrm{comp}}$,
\begin{equation}
  \langle \theta^2 \rangle \ \sim \ \frac{1}{\pi} 
  \frac{S^2}{U^2\ks^2}\frac{1}{\ku \delta}
  \left( \int_0^{\fracs{\pi}{2\ku \delta}} F(\eta)^2
  d\eta+\int_0^{\fracs{\pi}{2\ku \delta}} G(\eta)^2 \frac{\ku^2 \eta^2
  \delta^2}{\sin^2(\ku \eta \delta)} d\eta \right).
\end{equation}
Here we are interested in the scaling as $\delta\rightarrow 0$ so we are
justified in replacing the upper limit of the integral of $F(\eta)^2$ by
infinity.  Slightly more care must be taken with the integral involving
$G(\eta)^2$: because $\ku^2 \eta^2 \delta^2/\sin^2(\ku \eta \delta)$ is
uniformly bounded by $0$ and $\pi/2$ we can appeal to Lebesgue's dominated
convergence theorem to deduce
\begin{equation}
  \langle \theta^2 \rangle \ \sim \ \frac{1}{\pi}  \frac{S^2}{U^2\ks^2}
  \frac{1}{\ku\delta}
\left( \int_0^\infty F(\eta)^2 d \eta + \int_0^\infty G(\eta)^2 d \eta\right).
\end{equation}
The unstirred variance is $\langle \theta_0^2 \rangle=S^2/\kappa^2 \ks^4$, 
so the mixing efficiency is
\begin{equation}
  {\cal E}_{0} \ \sim \ \sqrt{\frac{\pi}{2}} \,
  \left(\int_0^\infty F(\eta)^2 d \eta + 
  \int_0^\infty G(\eta)^2 d \eta \right)^{-1/2} \, 
   r^{1/3} \, \Pec^{5/6}.
   \label{A0}
\end{equation}

\begin{figure}
\centerline{\includegraphics[width=10cm]{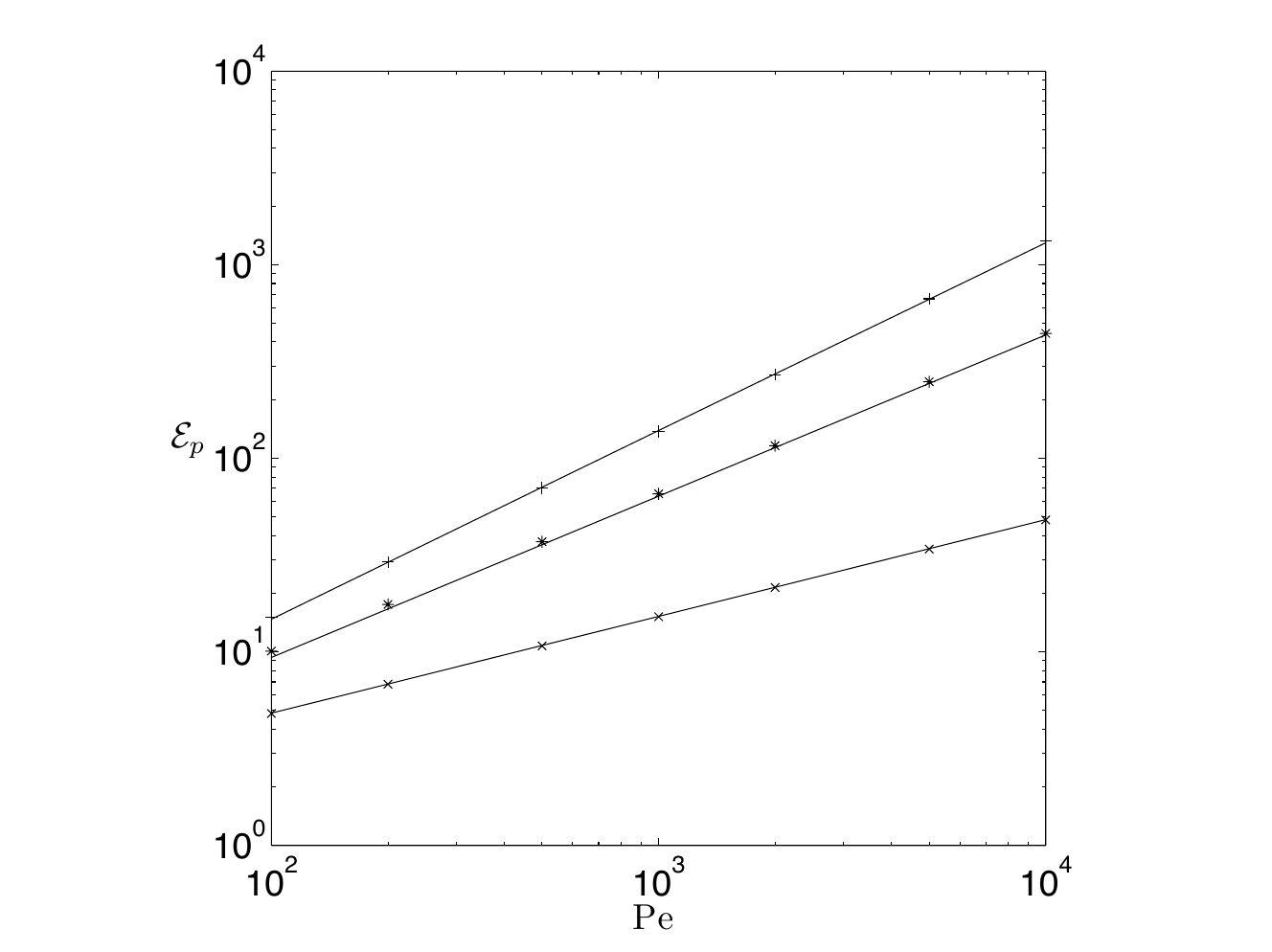}}
\setlength{\abovecaptionskip}{-3pt}
\caption{ Mixing efficiencies ${\cal E}_{p}$ for $p=1$ (x), $p=0$ (*) and
$p=-1$ (+) for the steady sine flow.  The solid lines are the scaling
predictions of the asymptotic analysis and the discrete data are from the
numerical solution of the steady advection--diffusion equation
(\ref{eq:ADsteady}).  }
\label{dumfig5}%
\end{figure}

Figure~\ref{dumfig5} shows ${\cal E}_{0}$ as a function of $\Pe$ from the
numerical solution of the steady advection--diffusion equation
(\ref{eq:ADsteady}).  The data accurately match the $\Pe^{5/6}$ scaling and
using the prefactor calculated from (\ref{A0}) there is only a 5\% discrepancy
between the internal layer solution and that calculated from the direct
numerical simulation at $\Pe=1000$ with $r=1$.  Remarkably, this $\Pe^{5/6}$
scaling also fits the data for the time-dependent stirring of the tilted
source shown in Figure~\ref{dumfig2}.

The gradient variance in the steady solution is given by
\begin{equation}
  \langle | \nabla \theta|^2 \big\rangle = \frac{\ks^2}{2} \left[ \langle f^2
  \rangle + \langle g^2 \rangle \right] +\frac{1}{2} \left[ \langle (f')^2
  \rangle + \langle (g')^2 \rangle \right].
  \label{eq:gradvarinternal}
\end{equation}
The $\langle f^2 \rangle$ and $\langle g^2 \rangle$ terms were computed above
in the context of ${\cal E}_{0}$, so we focus on the derivative terms.
Inserting the composite asymptotic solution~\eqref{eq:fginternalsol} we obtain
\begin{multline*}
  \frac{1}{2} \left[ \langle (f')^2 \rangle +  \langle (g')^2 \rangle
    \right]  \sim \left \langle \left( \frac{S}{U\ks} \frac{1}{\ku \delta^2}
  F'(y/\delta) \right)^2  \right \rangle + \\
  \left \langle \left(\frac{S}{U\ks} \frac{1}{\ku \delta^2} G'(y/\delta)
    \frac{\ku y}{\sin(py)} + \frac{S}{U\ks} \frac{1}{\ku\delta} G(y/\delta)
    \left[\frac{\ku y}{\sin(\ku y)}\right]' \right)^2 \right \rangle.
\end{multline*}
The last term is smaller (${\cal O}(\delta^{-1})$) than the immediately
preceeding term (${\cal O}(\delta^{-2})$) and may thus be neglected.  The
first two terms in~\eqref{eq:gradvarinternal} are also ${\cal
O}(\delta^{-1})$, so they may be neglected as well.  Hence
\begin{equation*}
  \frac{1}{2} \left[ \langle (f')^2 \rangle + \langle (g')^2 \rangle \right]
  \sim \frac{1}{\pi} \frac{S^2}{U^2 \ks^2} \frac{1}{\ku\delta^3}  \left[
  \int^{\fracs{\pi}{2\ku\delta}}_0 F'(\eta)^2 d\eta +
  \int^{\fracs{\pi}{2\ku\delta}}_0 G'(\eta)^2 \frac{\ku^2 \eta^2
  \delta^2}{\sin^2(\ku\eta\delta)}d\eta \right]
\end{equation*}
and by the same arguments as for the variance above we deduce
\begin{equation}
  \langle | \nabla \theta|^2 \big\rangle \sim \frac{1}{\pi} \left(
  \int_0^\infty F'(\eta)^{2} d \eta + \int_0^\infty G'(\eta)^{2} d \eta\right)
  \frac{S^2}{U^2\ks^2} \frac{1}{\ku \delta^3}\,.
\end{equation}
Using $\langle| \nabla \theta_0|^2 \rangle=S^2/\kappa^2 \ks^2$,
we deduce that the mixing efficiency is
\begin{equation}
  {\cal E}_{1} \ \sim \  \sqrt{\frac{\pi}{2}} \,
  \left( \int_0^\infty F'(\eta)^2 d \eta + \int_0^\infty
  G'(\eta)^2 d \eta \right)^{-1/2} \,
  \Pec^{1/2}.
  \label{A1}
\end{equation}

Figure~\ref{dumfig5} also shows the scaling of ${\cal E}_{1}$ from the 
numerical solution of (\ref{eq:ADsteady}). 
The internal layer asymptotic approximation in (\ref{A1}) and the numerical
solutions differ by approximately 1\% at $\Pe$ = 1000.  
The (lack of) scaling in $r$ was also confirmed numerically. 
Interestingly, for this problem stirring at ever smaller scales does {\em not} 
further enhance the mixing efficiency on small scales. 
This is apparently because the decrease in gradient variance due to stirring 
on small scales is compensated by the increase in 
gradient variance in the internal layer.

The inverse-gradient variance,
\begin{equation}
  \langle | \nabla^{-1} \theta|^2 \rangle \ = \ \langle | \nabla^{-1}(f(y)
  \sin(\ks x) + g(y) \cos(\ks x))|^2 \rangle,
\end{equation}
requires a slightly more subtle analysis.
Expanding $f(y)$ and $g(y)$ in Fourier series,
\begin{equation}
  f(y)=\sum _{n=0}^{\infty} f_{n} \cos(n\ku y) , \quad \quad
  g(y)=\sum _{n=1}^{\infty} g_{n} \sin(n\ku y)  
\end{equation}
the inverse gradient variance is
\begin{equation}
  \langle | \nabla^{-1} \theta|^2 \rangle \ = \ 
  \frac{f_{0}^{2}}{2 \ks^{2}} + 
  \frac{1}{4} \sum _{n=1}^{\infty} \frac{f_{n}^{2}+g_{n}^{2}}
  {\ks^{2} + n^{2} \ku^{2}}.
\end{equation}
We know that ${\cal E}_{-1} \lesssim \Pe$, so if we could establish a {\it lower} bound with 
the same scaling then we could conclude that ${\cal E}_{-1}$ scales classically $\sim \Pe$.
Hence we focus on deducing an asymptotic {\it upper} bound on 
$\langle | \nabla^{-1} \theta|^2 \rangle$.
As will be seen, however, the best we can do is assert the classical scaling 
with a logarithmic correction.

Toward that end, using the fact that $\sum_{n=1}^{\infty} n^{-2} = \pi^{2}/6$,
\begin{equation}
  \langle | \nabla^{-1} \theta|^2 \rangle \ \le \ 
  \frac{f_{0}^{2}}{2 \ks^{2}} \ + \ 
  \frac{\pi^{2}}{24 \ku^{2}} \, \sup_{n \ge 1} (f_{n}^{2}+g_{n}^{2}).
\end{equation}
Then the key is to note that the Fourier coefficients $f_{n}$ are all 
${\cal O}(S/U\ks)$---without any further factors of $\delta$ appearing---while 
the $g_{n}$ are ${\cal O}(S/U\ks \times \log{\frac{1}{\delta}})$.
Indeed,
\begin{equation}
  |f_{n}| \ = \ \left| \frac{(2-\delta_{n,0})\ku}{\pi}
  \int_{-\pi/2\ku}^{\pi/2\ku} f(y) \cos(n \ku y) dy \, \right| 
    \ \lesssim \ \frac{2(2-\delta_{n,0})S}{\pi U\ks}
    \int_{0}^{\infty} |F(\eta)|  \, d\eta
\end{equation}
while %**
\begin{eqnarray}
  |g_{n}| \ &=& \left| \frac{2\ku}{\pi}
  \int_{-\pi/2\ku}^{\pi/2\ku} g(y) \sin(n \ku y) dy \, \right|
    \nonumber \\
      &\sim& \frac{2S}{\pi U \ks} \left| \int_{-\pi/2\ku\delta}^{\pi/2\ku\delta} 
    \frac{\ku \delta \, \eta}{\sin(\ku \delta \, \eta)}
    G(\eta) \sin(n \ku \delta \, \eta) d\eta \, \right|
    \nonumber \\
      &\le& \frac{2 S}{U\ks} \int_{0}^{\pi/2\ku\delta} 
       | G(\eta)| \, d\eta \,  \ \sim \
       \frac{2S}{U\ks}\log\left[ \frac{1}{\ku \delta}\right].
\end{eqnarray}
In the last step above we used the fact that $|G(\eta)| \sim \eta^{-1}$ as 
$\eta \rightarrow \infty$.

The unstirred large-scale variance is $\langle| \nabla^{-1} \theta_0|^2
\rangle=S^2/\kappa^2 \ks^6$, so we deduce the large-scale 
mixing efficiency obeys
\begin{equation}
  {\cal E}_{-1} \ \ge \ C \, \frac{\Pe}{\ks L}
  \times \frac{1}{\sqrt{{1 + C' \left(\frac{\log{\Pe}}{r}\right)^{2}}}}
\end{equation}
where $C$ and $C'$ are absolute constants.  The numerical solutions of the
steady advection--diffusion equation shown in Figure~\ref{dumfig5} confirms
this classical scaling; the logarithmic term is not visible for this range of
$\Pe$.  Again, this is consistent with the dynamic data for ${\cal E}_{-1}$ in
Figure~\ref{dumfig2}.
%*%*%*

%===================================================

%\newpage

\section{Summary and Discussion}
\label{sec:SandD}

\noindent

The suppression of bulk variance of a scalar field is a natural gauge of the
efficiency of stirring, and this notion allows for the examination of the
effect of a flow on various length scales.  We have quantified the influence
of stirring on weighted bulk variances in terms of nondimensional mixing
efficiencies ${\cal E}_p$ that characterize fluctuations on relatively small
($p=1$), intermediate ($p=0$) and large ($p=-1$) length scales.  We studied
these mixing efficiencies for statistically stationary, homogeneous and
isotropic flow fields stirring scalars sustained by a variety of steady but
spatially inhomogeneous scalar sources and sinks on periodic domains.  We
reach a number of conclusions:
\begin{itemize}

\item{Very generally, the mixing efficiencies ${\cal E}_p$ are bounded from
below $\sim \Pe^0$ at high P\'eclet numbers---indicating ineffective
stirring---and from above $\sim \Pe^1$---the classical scaling anticipated by
the simplest eddy diffusion theory.  Classical scaling of the efficiencies
corresponds to the existence of residual supression of variance in the
vanishing diffusion ($\kappa \rightarrow 0$) limit.}
\item{Source, sink and statistically stationary homogeneous
and isotropic flow combinations exist that realize both the upper and lower
bounds' scaling with respect to $\Pe$---and in some cases the prefactors
as well---showing that the analysis is sharp at the most general level.}

\item{The small and intermediate scale mixing efficiencies
${\cal E}_1$ and ${\cal E}_0$ are ultimately limited
by length scales in the source-sink distributions.
That is, even if the efficiencies scale classically the inferred mixing lengths
$\sim \kappa \Pe/U$ are limited by length scales defined by the
sources and sinks rather than by the spectrum of scales in the flow.
On the other hand the upper bound analysis does not prevent small scales 
in the flow field from enhancing suppression of the variance at large scales, 
i.e., ${\cal E}_{-1}$, to an unlimited degree.}

\item{Sufficiently ``rough'' scalar sources and sinks, in particular
non-square-integrable distributions, necessarily produce sub-classical scaling
of ${\cal E}_0$ and/or ${\cal E}_1$ at high P\'eclet numbers.  In such cases
there is no residual suppression of variance in the vanishing diffusion limit
for any finite mean kinetic energy statistically stationary homogeneous and
isotropic stirring.}

\item{The mixing efficiencies at various length scales need not scale the same
as $\Pe \rightarrow \infty$.  This was illustrated explicitly by the example of a monochromatic
source distribution stirred by the random sine flow.  We found by direct
numerical simulation and an adiabatic asymptotic internal layer analysis that
for small scales ${\cal E}_1 \sim \Pe^{1/2}$, for intermediate scales ${\cal
E}_0 \sim \Pe^{5/6}$, and the large length scale efficiency ${\cal E}_{-1}$
exhibits near-classical scaling $\sim \Pe/\log{\Pe}$.}
\end{itemize}

The analysis, simulations and modeling presented in this paper have addressed
some fundamental questions regarding suppression of the long time averaged
bulk scalar variance via stirring by incompressible and statistically
homogeneous and isotropic flow fields, but compelling challenges remain for
future study.  Among the unsolved problems open questions are:
\begin{itemize}
\item{Can the anomalous mixing efficiency bounds for measure-valued or fractal
sources be achieved by any statistically stationary homogeneous and isotropic
flow?  This is a difficult problem for direct numerical simulations of the
advection--diffusion equation because of the small spatial scales that need to
be resolved.  An alternate simulation approach is particle tracking, by
following the evolution of many discrete particles moving with the flow and
diffusing and approximating the continuous scalar concentration on an
appropriately coarse grained level.}

\item{Can the small-flow-scale enhancement of the large-scale mixing
efficiency ${\cal E}_{-1}$, as suggested by the upper bound in
(\ref{eq:igvarlubound}), be realized?  It makes sense that stirring may be
capable of suppressing large-scale variance, even for negligible molecular
diffusion, by transferring scalar inhomogeneities from large length scales to small
length scales via kinetic stretching and folding mechanisms.  The upper bound
provides a quantitative estimate of this effect and it will be interesting to
see to what extent it is an accurate estimate.}

\item{What is the mixing efficiency of ``real'' stationary homogeneous and
isotropic turbulence in two and three dimensions?  The bounds derived here all
apply to these flows, and the inevitable limitations on the mixing efficiency
imposed by the source and sink structure apply. Turbulent mixing by a fluid
with viscosity $\nu$ is often characterized by a Reynolds number
$\text{Re}=UL/\nu$ and a Schmidt number $\text{Sc}=\nu/\kappa$.  While the
mixing efficiencies ${\cal E}_p$ are studied here as functions of
$\Pe=\text{Re} \times \text{Sc}$, the separate Reynolds and Schmidt number
dependences are of interest, too.  These questions can be investigated
theoretically and via direct numerical simulations in the periodic domain
setting utilized here.}

\item{The uniform-wind optimal solution presented in
Section~\ref{sec:saturation} is only appropriate for a one-dimensional source
distribution in a domain with periodic boundary conditions.  
It is optimal by virtue of always having a component blowing
along the source gradient, so no time and little energy
is wasted blowing along lines of constant source.
If the source has a more complicated structure, then it is not generally possible to
achieve such an efficient flow while preserving incompressibility.  Physical
boundaries or a different golbal topology (e.g. a spherical domain as relevant to
geophysical applications) will likewise invalidate the uniform-wind solution.
The general question is then: For a given distribution of sources and sinks in
a given domain, what is the optimal stirring strategy to achieve the greatest
reduction of bulk variance?  This question has obvious relevance to
engineering applications.  While we have answered it in the simplest setting
of a single mode source-sink distribution on a periodic domain, it is clearly
a complex problem in general.  It is not unlikely, however, that intuition
gained from the study of idealized models may contribute to the development of
valuable intuition for such systems.}

\item{Bounded domains with rigid walls are more appropriate for some
engineering applications.  Then the scalar sources and/or sinks can be
implemented by boundary conditions rather than as body sources and sinks
\cite{Balmforth2003}.  We can also imagine situations where fluid and/or
imposed scalar fluxes at boundaries contribute to the scalar variance in the
bulk.  How this will affect the efficiency scalings is an open question.}

\item{We have seen that very generally ${\cal E}_1 \ge 1$, but have only
established a similar lower bound on ${\cal E}_0$ and ${\cal E}_{-1}$ for the
simplest case of monchromatic sources.  It is an open question whether or not
there exist source, sink and flow (even steady flow) combinations where ${\cal
E}_0$ and/or ${\cal E}_{-1} < 1$.  It is worthwhile noting in this context
that in all cases $\langle \theta^2 \rangle^2 \le \langle |\nabla \theta|^2
\rangle \langle |\nabla^{-1}\theta|^2 \rangle$.  Any stirring decreases the
mean bulk gradient variance, but this suggests that if it does not decrease
the variance significantly then it would have to {\it increase} the inverse
gradient variance.  Hence it may not be surprising to find flows that can
amplify large-scale fluctuations.}

\item{A related problem of interest is the advection--diffusion of a passive
scalar $\theta$ with decay rate $\zeta$ sustained by a body
source with evolution described by
\begin{equation}
  \frac{\partial \theta}{\partial t} + \bm{u} \cdot \nabla \theta= \kappa
  \lapl \theta +s(\bm{x}) - \zeta\theta.
\end{equation}
In applications $\zeta$ may have an interpretation as a chemical reaction rate
or a radiative relaxation rate relevant in meteorology or for other
geophysical flows on the sphere.  A linear amplitude damping term in the
advection--diffusion equation introduces new features such as competition
between scalar decay and mixing to supress steady-state variances
\cite{ShawGFD2005}.}

\item{Finally, all the questions above can be posed for time dependent 
sources and sinks.}
\end{itemize}

%\noindent
%``It has been recognized that $L^p$ norms of the passive scalar fail to quantify
%the \emph{stirring efficiency} of a mixing process because they are
%insensitive to small-scale structures \cite{Danckwerts1952}.''

\section*{Acknowledgements}

We are grateful for stimulating discussions with Paola Cessi, Bruno
Eckhardt, Paul Federbush, Matthew Finn, Francesco Paparella, Grigorios
Pavliotis, J\"org Schumacher, William R. Young, and many of the
participants of the 2005 GFD Program at Woods Hole Oceanographic
Institution where much of this work was performed.  TAS was supported
in part by the National Science and Engineering Research Council of
Canada through a Canadian Graduate Scholarship.  J-LT was supported in
part by the UK Engineering and Physical Sciences Research Council
grant GR/S72931/01.  CRD was supported in part by NSF grants
PHY-0244859, PHY-0555324 and an Alexander von Humboldt Research Award.

%\newpage

\appendix

\section{Lower Bound on Variance Efficiency}
\label{sec:lowbound0}

One might expect the variance efficiency ${\cal E}_0$ to have a lower bound of
unity, implying that stirring \emph{always} decreases the variance.  In order
to derive a uniform (in $\Pe$) lower estimate we can search for the maximum
variance subject to the steady-state production-dissipation constraint:
\begin{equation}
  \langle \theta^2 \rangle \geq \min_{\vartheta}~ \{ \langle
  \vartheta^2 \rangle~|~ \kappa \langle | \nabla \vartheta |^2
  \rangle = \langle s \vartheta \rangle \} .
\end{equation}
The solution of this optimization problem in Fourier space is
\begin{equation}
  \hat{\theta}(\bm{k})= \frac{\mu}{2} \frac{\hat{s}(\bm{k})}{\mu \kappa k^2
  + 1}
\end{equation}
where $\mu$ is the Lagrange multiplier enforcing the constraint
\begin{equation}
  \kappa \sum_{\bm{k}} k^2 | \hat{\theta}(\bm{k})|^2 =
  \sum_{\bm{k}} \hat{\theta}(\bm{k}) \hat{s}^*(\bm{k})~~
  \quad \Longrightarrow \quad \kappa \sum_{\bm{k}} \frac{\mu k^2
  |\hat{s}(\bm{k})|^2}{(\mu \kappa k^2 +1)^2} = 2 \sum_{\bm{k}}
  \frac{|\hat{s}(\bm{k})|^2}{\mu \kappa k^2 +1}.
\label{sums}
\end{equation}
where $*$ denotes the complex conjugate.

In the case of a dichromatic source with wavenumber $k_1$ at amplitude $s_1$
and $k_2$ at amplitude $s_2$, the constraint requires one to solve a cubic
equation for $\xi=1/\mu \kappa k_1^{2}$,
\begin{equation}
  (1+\alpha)\xi^3 + \tfrac{1}{2} (1+\alpha \beta +4 \beta +4 \alpha) \xi^2 +
  (\beta +\alpha \beta + \beta^2 +\alpha) \xi +\tfrac{1}{2} (\beta^2+\alpha
  \beta)=0
\end{equation}
where $\alpha=s_1^2/s_2^2$ and $\beta= k_1^2/k_2^2$.
Then the mixing efficiency bound is
\begin{equation}
  {\cal E}_0^2 \ge \frac{\sum_{k_{1}, k_{2}}|\hat{\theta}_0(\bm{k})|^2}
  {\sum_{k_{1}, k_{2}}|\hat{\theta}(\bm{k})|^2} =
  \frac{4 (1+ \frac{\alpha}{\beta^2})}
  {\frac{1}{(1+\xi)^2}+\frac{\alpha}{(\beta+\xi)^2}},
\end{equation}
and a numerical evaluation of the roots reveals that the minimum value of the
efficiency bound is less than 1 for $\forall~\xi$.  Thus the variance
production-dissipation balance alone does {\it not} rule out the existence of
flows that could \emph{increase} the scalar variance.

%\newpage

\section{Steady uniform wind on a $\delta$-function source}

Taking the Fourier transform of the steady advection--diffusion
equation~\eqref{eq:AD} with a uniform velocity field along the $x_{d}$-axis,
\begin{equation}
  %\sum_k
  \hat{\theta}(\bm{k}) =
  %\sum_k
  \frac{\hat{s}(\bm{k})}{\kappa k^2 + i U k_d}
\end{equation}
where $k_d$ is the $d$th component of the horizontal wavenumber.
Now substitute~$\hat{s}(\bm{k})=\text{const}$ for a~$\delta$-function
source and after approximating the sums by integrals (since we are only
interested in the asymptotic behavior), we find
\begin{subequations}
\begin{alignat}{2}
  d&=2: \qquad &\langle |\nabla^{p}\theta|^2 \rangle &= \int_0^{2\pi} d\phi
  \int_{2\pi/L}^{\infty} \frac{k^{2p+1} dk}{\kappa^2 k^4 + U^2
    k^2\cos^2\phi}\,,\\
  d&=3: &\langle| \nabla^{p}\theta|^2 \rangle &= \int_0^{2\pi} d\phi
  \int_0^{\pi} \sin \vartheta d\vartheta \int_{2\pi/L}^{\infty} \frac{k^{2(p+1)}
  dk}{\kappa^2 k^4 + U^2 k^2\cos^2\vartheta}\,.
\end{alignat}%
\label{eq:integrals}%
\end{subequations}%
The variances in the absence of stirring are found by calculating the
integrals~\eqref{eq:integrals} with $U=0$. Straightforward evaluation of the
integrals~\eqref{eq:integrals} yields
\begin{subequations}
\begin{alignat}{4}
  d&=2:
  \qquad
  &{\cal E}_{1} &= 1,
  \quad &{\cal E}_{0} &\sim \sqrt{\Pe}\,,
  \quad &{\cal E}_{-1} &\sim \sqrt{\Pe}\,,\\
  d&=3:
  &{\cal E}_{1} &= 1,
  \quad &{\cal E}_{0} &\sim 
  \frac{\sqrt{\Pe}}{\sqrt{\log \Pe}}\,,
  \quad &{\cal E}_{-1} &\sim \sqrt{\Pe}\,.
\end{alignat}
\end{subequations}
These anomalous scalings in $\Pe$ suggest that the uniform flow is not the
optimal allowed by the bound for the $\delta$-function source in both $d=2$
and~$3$. This emphasizes the source-dependent nature of the optimal stirrer.

%\newpage

\section{Steady shear at an angle}

\begin{figure}
\centerline{\includegraphics[width=15cm]{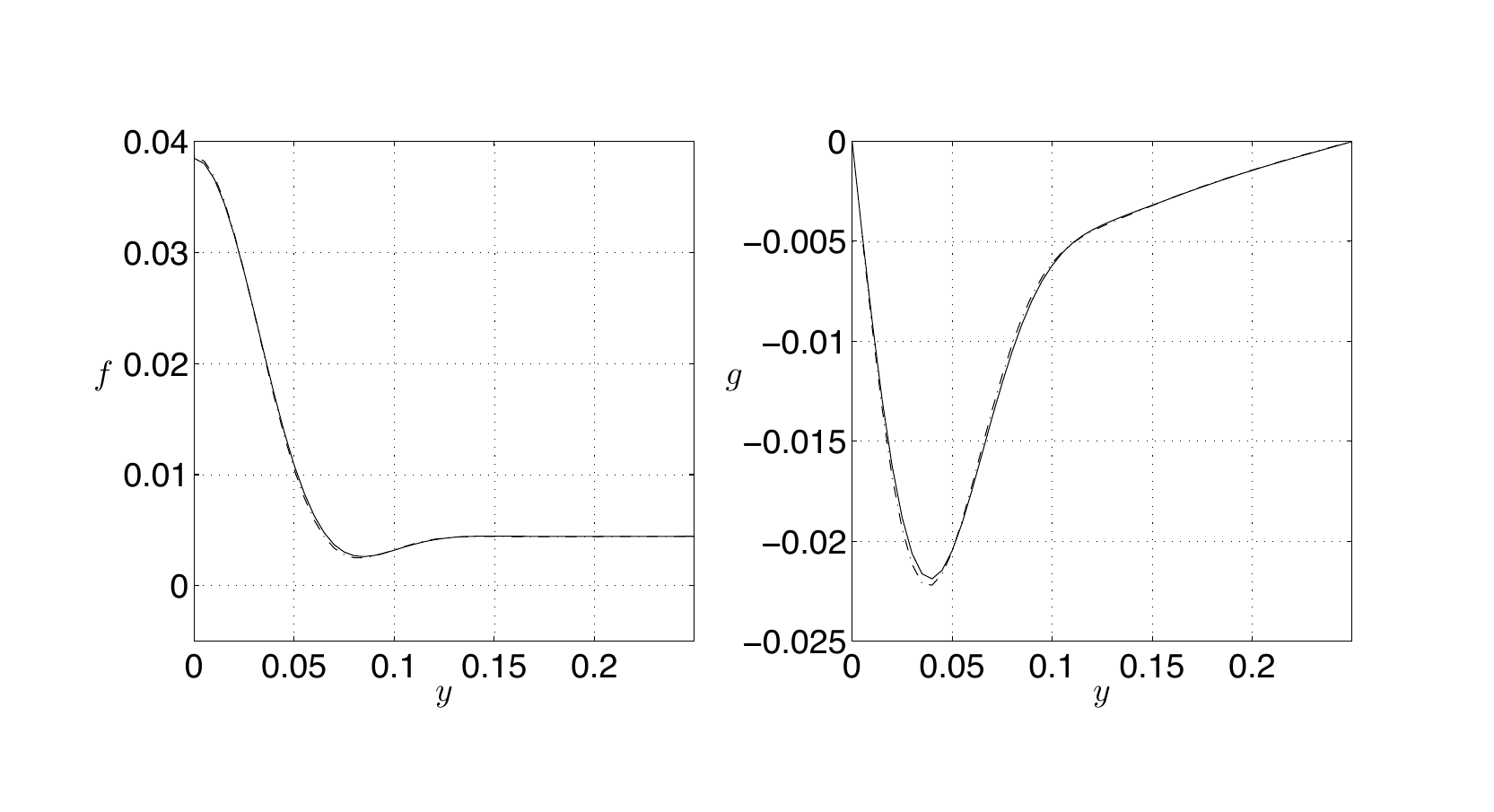}}
\setlength{\abovecaptionskip}{-30pt}
\caption{Comparison of the direct numerical solution (solid) and the internal
layer solution (dashed) for \hbox{$\Pe$ = 1000}.}
\label{dumfig6}
\end{figure}

\begin{figure}
\centerline{\includegraphics[width=20cm]{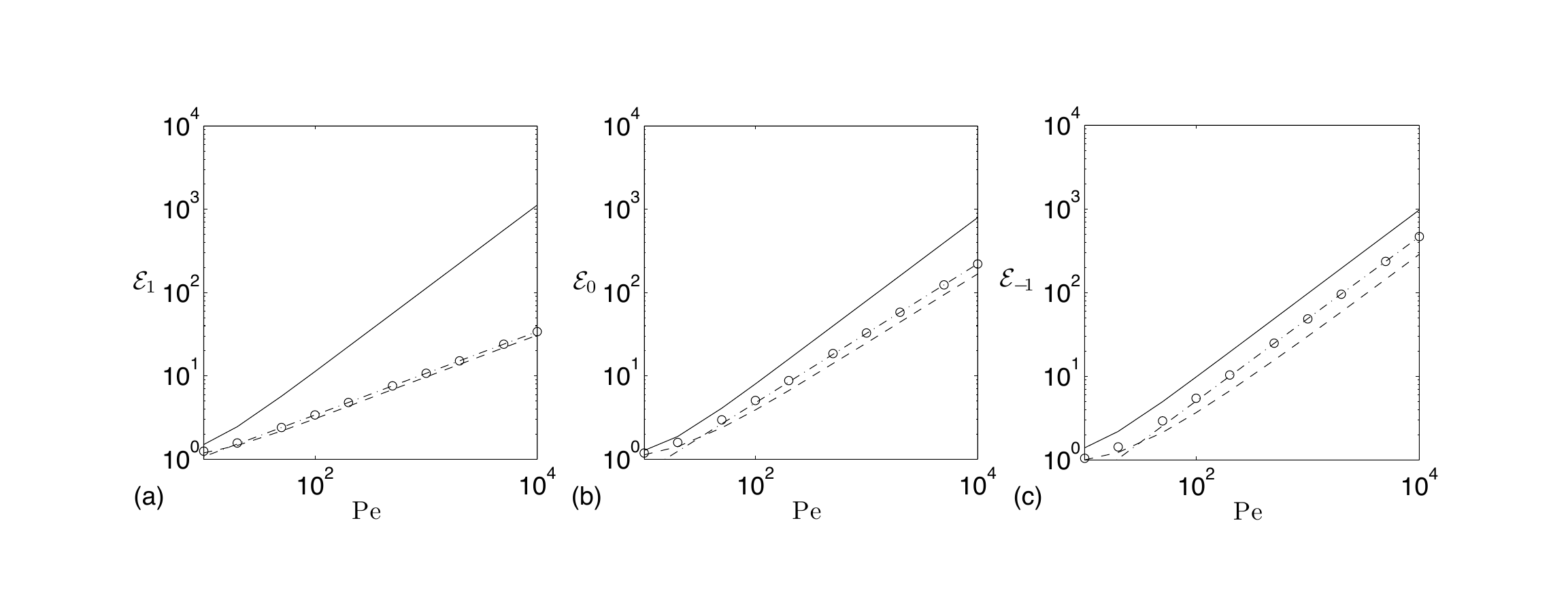}}
\setlength{\abovecaptionskip}{-20pt}
\caption{Mixing efficiencies ${\cal E}_p$ as a function of $\Pe$ for (a)
$p=1$, (b) $p=0$, and (c) $p=-1$. The solid lines are the bounds, the
dashed-dot lines are scalings predicted from the internal layer asymptotic
analysis, the open circles are the discrete data from the numerical solution
of the steady advection--diffusion equation (\ref{eq:ADsteady}), and the
dashed lines are the result of the direct numerical solution for the
time-dependent random sine flow.}
\label{dumfig7}%
\end{figure}

When the source is tilted at a 45$^{\circ}$ angle, i.e. $s(\bm{x}) = \sqrt{2} S \sin(k_s(x+y))$, the functions $f$
and $g$ must satisfy
\begin{subequations}
\begin{align}
  -\sqrt{2}U\ks \sin(\ku y)\, g(y)
  &= \kappa \left [ -\ks^2 + \frac{d^2}{dy^2}
    \right] f(y) + \sqrt{2} S\sin(k_s y)\,,\label{B1a}\\
  \sqrt{2}U\ks \sin(\ku y) f(y) &= \kappa \left [ -\ks^2 + \frac{d^2}{dy^2}
    \right] g(y)+ \sqrt{2}S \cos(k_s y)\,.\label{B1b}
\end{align}
\label{eq:fgeqnstilted}%
\end{subequations}%
Upon changing variables as done in section $\rm{VII}$ we obtain
\begin{subequations}
\begin{align}
  \frac{1}{\Pec}  \left[ -1 + \frac{d^2}{d \tilde{y}^2} \right]
  \hat{f}(\tilde{y})+\sin \tilde{y} &= -\sin(r\tilde y) \hat{g}(\tilde{y}),\label{B2a}\\
  \frac{1}{\Pec}  \left[ -1 + \frac{d^2}{d \tilde{y}^2} \right]
  \hat{g}(\tilde{y}) + \cos \tilde{y} &= \sin(r\tilde y) \hat{f}(\tilde{y}).\label{B2b}
\end{align}
\label{eq:fgeqnsscaledtilted}%
\end{subequations}%
The internal layer solution is once again obtained by expanding in powers of
$\Pec^{-1}$. To leading order the solution to~\eqref{eq:fgeqnsscaledtilted} in 
the outer region is
\begin{equation}
  \hat{f}_{\mathrm{outer}} \sim 1,\qquad 
  \hat{g}_{\mathrm{outer}} \sim -\frac{1}{ \tan(r \tilde{y})}\,.
  \label{eq:outersoltilted}
\end{equation}
The inner internal layer solution is the same as for the untilted source
 problem.  In Figure~\ref{dumfig6} we compare the internal layer solution
 against the exact (numerical) solution of the full advection--diffusion
 equation (\ref{eq:ADsteady}) at $\Pe=1000$ where the small parameter
 $\epsilon \approx .2$.
Recovering all the scalings and
letting $\delta=\epsilon/r^{1/3} \ks$, we define the composite approximations
\begin{subequations}
\begin{align}
  f_{\mathrm{comp}}(y) \ &= \ \frac{S}{U\ks} \,[\hat{f}_{\mathrm{inner}}(y) + f_{\mathrm{outer}}(y) ]
  %= \frac{S}{U\ks}
  %\frac{r^{-2/3}}{\epsilon}F \left( \frac{r^{1/3} \ks}{\epsilon} y \right)
  \ = \frac{S}{U\ks}\, \left[ \frac{1}{\ku \delta} F \left(\frac{y}{\delta}\right)+1\right],\\
  g_{\mathrm{comp}}(y) \ &= \ \frac{S}{U\ks} \,
  \ku y \times \hat{g}_{\mathrm{inner}}(y) \times \hat{g}_{\mathrm{outer}}(y)
  %=\frac{S}{U\ks} \frac{r^{-2/3}}{\epsilon}G
  %\left( \frac{r^{1/3} \ks}{\epsilon} y \right) \frac{\ku y}{\sin(\ku y)}
  %\\
  \ = \frac{S}{U\ks}\, \frac{1}{\ku \delta}\,
  \frac{\ku y}{\tan(\ku y)}\,G\left(\frac{y}{\delta}\right).
\end{align}
\label{eq:fginternalsoltilted}%
\end{subequations}%
Given the internal layer solution we can compute the multi-scale mixing
measures.  In fact, to leading order the dependence of the efficiencies on the
P\'eclet number is the same. The difference is in the prefactors. Figure
\ref{dumfig7} compares the scalings derived from the internal layer solution
(including prefactors) to the discrete data from the numerical solution and
the numerical solution using the random sine flow.

\section{Corrigendum (29 April 2011)}
\label{sec:corr}

%
% Math symbols
%
\renewcommand{\l}{\left}            % \left
\renewcommand{\r}{\right}           % \right

\mathnotation{\xc}{x}               % Position (component)
\mathnotation{\xv}{\bm{\xc}}        % Position (vector)
\mathnotation{\kc}{k}               % Wavevector (component)
\mathnotation{\kv}{{\bm{\kc}}}      % Wavevector (vector)
\mathnotation{\Lsc}{L}              % Periodic length
\mathnotation{\Eff}{\mathcal E}     % Mixing efficiency
\mathnotation{\src}{s}              % Source
\mathnotation{\K}{K}                % Gravest wavenumber
\mathnotation{\mono}{\text{mono}}   % monochromatic prefix
\mathnotation{\kcs}{\kc_{\mathrm{s}}}% monochromatic source wavenumber
\mathnotation{\smiley}{F}

In this corrigendum we rectify an error in the published version of
this preprint (Shaw et al.~\cite{Shaw2007}).  In
Section~\ref{sec:bound_{lower}} we obtained an estimate on the
variance of the concentration~$\theta(\xv)$ of an advected passive
scalar.  We did this by solving the optimization problem
\begin{equation}
  \langle \theta^2 \rangle \le \max_{\vartheta} \,
  \{ \langle \vartheta^2 \rangle \, | \,
  \kappa \langle |\nabla \vartheta|^2 \rangle = 
  \langle s \vartheta \rangle
  \}
  \label{eq:H0max}
\end{equation}
where the angle brackets denote spatial averaging, $\kappa$ is the
diffusivity, and~$\src(\xv)$ is the spatially-dependent source with
spatial mean zero.  The constraint in~\eqref{eq:H0max} arises from
integrating the advection-diffusion equation with periodic boundary
conditions.  The Euler--Lagrange equation for the extremizer
$\vartheta_{*}(\xv)$ is
\begin{equation}
  2 \vartheta_{*} - 2\mu \kappa \Delta \vartheta_{*} - \mu s = 0
  \label{eq:H0maxEL}
\end{equation}
where~$\mu$ is the Lagrange multiplier enforcing the constraint
in~\eqref{eq:H0max}.  In terms of the Fourier coefficients the
solution of~\eqref{eq:H0maxEL} is straightforward:
\begin{equation}
  {\hat{\vartheta}_*}{}_\kv = \tfrac12{\mu} \, \frac{\hat{\src}_\kv}
  {\mu \kappa \kc^2 + 1},
  \qquad
  \sum_{\kv} \frac{2 + \mu \kappa \kc^2}{(\mu \kappa \kc^2 + 1)^2}\,
    |\hat{\src}_\kv|^2 = 0.
  \label{eq:H0maxELsol}
\end{equation}
that latter being an equation for $\mu < 0$.  If the source is
``monochromatic,'' i.e., is an eigenmode of the Laplacian with
eigenvalue~$-\kcs^2$, then the solution~\eqref{eq:H0maxELsol}
simplifies to
\begin{equation}
  \mu^\mono = -2/(\kappa \kcs^2),
  \qquad \vartheta_*^\mono = \src/(\kappa \kcs^2)
  \qquad \text{(monochromatic source)}.
  \label{eq:monosol}
\end{equation}
From~\eqref{eq:monosol} Shaw et al.\ inferred that for monochromatic
sources the variance~$\langle \theta^2 \rangle$ could never decrease
below its value in the absence of any stirring flow.  That is, for
monochromatic sources there are no `unmixing' velocity fields which
raise variance rather than lower it.

The problem with the solution~\eqref{eq:H0maxELsol} (and its
monochromatic limit~\eqref{eq:monosol}) is that it is not always a
maximum for \eqref{eq:H0max}.  To see this,
let~$\theta=\vartheta_*+\delta\theta$.  We have
\begin{equation}
  \langle\theta^2\rangle = \langle\vartheta_*^2\rangle + \smiley,\qquad
  \smiley \ldef 2\langle\vartheta_*\delta\theta\rangle
  + \langle(\delta\theta)^2\rangle,
\end{equation}
so we want to investigate the sign of~$\smiley$ to determine
if~$\vartheta_*$ is indeed a maximum.  After enforcing the constraint
in~\eqref{eq:H0max} and integrating by parts we can derive
\begin{equation}
  \smiley
  = \mu\kappa\langle\lvert\grad\delta\theta\rvert^2\rangle
  + \langle(\delta\theta)^2\rangle.
  \label{eq:smileynearmax}
\end{equation}
from which an application of Poincar\'e's inequality gives
\begin{equation}
  \smiley
  \le \mu\kappa\langle\lvert\grad\delta\theta\rvert^2\rangle
  + \K^{-2}\langle\lvert\grad\delta\theta\rvert^2\rangle
  = \l(\K^{-2} + \mu\kappa\r)\langle\lvert\grad\delta\theta\rvert^2\rangle
\end{equation}
where~$\K=2\pi/\Lsc$ is the smallest allowable wavenumber, with~$\Lsc$
the size of the periodic domain.  Thus, $\smiley$ is guaranteed to be
negative (and~$\vartheta_*$ is a maximum) if
\begin{equation}
  -\mu\kappa\K^2  \ge 1.
  \label{eq:maxcriterion}
\end{equation}
(Showing that this is true when equality holds
in~\eqref{eq:maxcriterion} requires a slightly longer analysis.)  We
can also show that if~\eqref{eq:maxcriterion} is violated, then there
are values of~$\delta\theta$ that make~$\smiley$ negative, showing
that~\eqref{eq:maxcriterion} is necessary in addition to being
sufficient.

For monochromatic sources, inserting~\eqref{eq:monosol}
into~\eqref{eq:maxcriterion} implies that we are guaranteed to have a
maximum if
\begin{equation}
  \kcs^2 \le 2\K^2
  \qquad \text{(monochromatic source)}.
  \label{eq:monomax}
\end{equation}
Larger values of~$\kcs$ leave open the possibility of~$\vartheta_*$
not being a maximum, and indeed this is the case, as shown
in~\cite{Thiffeault2011_review}.  Thus, it is not true as claimed
in~\cite{Shaw2007} that all monochromatic sources
have~$\langle\vartheta_*^2\rangle$ as a maximum variance: the only
ones that do have small wavenumber~$\kcs$
satisfying~\eqref{eq:monomax}.  In particular this means that many
monochromatic sources can have mixing efficiency~$\Eff_0$ less than
unity (the same is true for~$\Eff_{-1}$; see~\cite{Shaw2007} for
definitions).

We emphasize that we expect that generic velocity fields will decrease
variance, and so will be `efficient' stirrers.  It is still an open
question to characterize the sources that cannot lead to an increase
in variance, that is, for which unmixing flows do not exist.

%\newpage
%\bibliography{journals_abbrev,greg,articles}

\end{document}